# Relativistic many-body calculations of multipole (E1, M1, E2, M2) transition properties in Al II

Yuan-Fei Wei[1,2,3], Zhi-Ming Tang[4,*], Xue-Ren Huang[3,5,*], Ming-Lu Bu[1], Xin-Ye Xu[4], Yi-Yu Cai[1,*]

1.  Macao Institute of Materials Science and Engineering (MIMSE), Faculty of Innovation Engineering, Macau University of Science and Technology, Taipa, Macau SAR 999078, P. R. China.
2.  Key Laboratory of Atom Frequency Standards, Innovation Academy for Precision Measurement Science and Technology, Chinese Academy of Sciences, Wuhan 430071, P. R. China.
3.  University of Chinese Academy of Sciences, Beijing 100049, P. R. China.
4.  State Key Laboratory of Precision Spectroscopy, East China Normal University, Shanghai 200241, P. R. China.
5.  Wuhan Institute of Quantum Technology, Wuhan 430206, P. R. China

## Abstract:

We present systematic relativistic many-body calculations of multipole transition properties for singly charged aluminum ion (Al II) using a method that combines the configuration interaction and many-body perturbation theory (CI+MBPT). Our calculations cover the 103 lowest energy levels in Al II. For five key low-lying configurations ($3s^2$ $^1S_0$, $3s3p$ $^3P_0$, $3s3p$ $^3P_1$, $3s3p$ $^3P_2$, $3s3p$ $^1P_1$), we tabulate the transition wavelengths, reduced matrix elements, transition probabilities, and oscillator strengths for about 400 electric dipole (E1), magnetic dipole (M1), electric quadrupole (E2), and magnetic quadrupole (M2) transitions arising from these levels. Our calculated values agree well with available experimental data and other high-precision theoretical calculations, with typical deviations on the order of 1%. Notably, we report over 80% of these transition lines as previously unreported, significantly expanding the existing spectroscopic database for Al II. These results can serve as a valuable reference resource for ongoing precision quantum metrology as well as astrophysical spectroscopy involving Al II ion.

Key words：Al II；Multipole transition properties；CI+MBPT；relativistic many-body calculations；Atomic structure calculation

* Corresponding author.
  Email address: zmtang@lps.ecnu.edu.cn; hxueren@wipm.ac.cn; yycai@must.edu.mo

# Contents

# 1. Introduction:

The structure of Group IIIA element singly charged ions can be simplified as a tightly bound atomic core with two outer valence electrons. Transitions between states with angular momentum quantum numbers J= 0 in these ions are highly insensitive to fluctuations in electromagnetic fields or other environmental perturbations. This property makes these ions, such as Al II, B II and In II, excellent candidates for optical frequency standards [1-5]. The Al II optical clock, in particular, has achieved exceptional precision with uncertainties of $10^{-19}$ magnitude [6]. The uncertainties of external field frequency shifts, caused by the fluctuations in the electromagnetic field or the imprecise understanding of atomic structure, represent significant contributions to the uncertainty of the Al II optical clock [6-11]. Accurate theoretical calculations of the atomic spectral properties of Al II, including the energies of states, the oscillator strength of the transitions between states, and polarizabilities, are therefore crucial in the advancement of both atomic theory and the execution of diverse targeted high-precision experiments, particularly in evaluating of external field frequency shifts for Al II optical clock. Precise theoretical modeling of atomic properties for Al II provides valuable reference data for astrophysical spectroscopy and helps illuminate cosmic evolution processes [12-14].

Previous theoretical and experimental studies have provided valuable insights into atomic properties for Al II. On the theoretical front, several computational approaches have been employed, including relativistic configuration interaction Dirac-Fock (CIDF) calculations [15], multi-configuration Hartree-Fock (MCHF) calculations [15,16], configuration interaction plus core polarization method (CICP) calculations [17], configuration-interaction plus coupled-cluster method (CI + all order) calculation [2], relativistic coupled-cluster (RCC) calculation [18], Fock-space relativistic coupled-cluster (FSRCC) calculation [19], and finite-field method (FF) calculation [20]. And the experimental studies mainly focus on the measurements of partial transitions' properties, such as, the absolute frequency and polarizabilities [1,6,8,11,21].

Early studies established the $3s^2$ $^1S_0$-$3s3p$ $^3P_0$ transition in Al II as an optical clock transition with exceptional precision [1–4, 6–11], and recent work has identified the $3s^2$ $^1S_0$-$3s3p$ $^3P_2$ transition as a promising candidate for both optical clock and quantum sensing to search for new physics [22]. However, prior investigations have predominantly centered on lower-energy states and electric dipole transitions, leaving critical gaps in the systematic characterization of higher-energy states and other transitions-such as the logic ($3s^2$ $^1S_0$-$3s3p$ $^3P_1$), theoretical cooling ($3s^2$ $^1S_0$-$3s3p$ $^1P_1$), the new clock transition with potential ($3s^2$ $^1S_0$-$3s3p$ $^3P_2$), and other multipole transitions essential for precision experiments [8]. A comprehensive study of multipole transition properties (e.g., wavelengths and amplitudes) is therefore vital to advance both fundamental understanding and applications of Al II.

In this work, we employ configuration-interaction and many-body perturbation theory (CI+MBPT) to perform a comprehensive atomic structure analysis for Al II. Our calculations determine the energies of the 103 lowest-lying states, comprising 58 even-parity and 45 odd-parity states. For the five lowest energy states ($3s^2$ $^1S_0$, $3s3p$ $^3P_0$, $3s3p$ $^3P_1$, $3s3p$ $^3P_2$, $3s3p$ $^1P_1$), we compute multipole transition properties including electric dipole (E1), electric quadrupole (E2), magnetic dipole (M1), and magnetic quadrupole (M2) reduced matrix elements, along with their associated transition probabilities and oscillator strengths. These states are particularly relevant for precision measurement applications with Al II, including optical atomic clocks, quantum logic spectroscopy, and quantum state manipulation experiments.

The remainder of this paper is organized as follows. Section 2 presents the theoretical framework, including general formulas for radiative transition properties and an overview of the CI+MBPT method as applied to two-electron systems. In Section 3, we compare our numerical results with available theoretical and experimental data, including reference values from the National Institute of Standards and Technology (NIST) atomic spectra database [23]. Section 4 concludes with a summary of our findings.

## 2. Method of Calculations

### 2.1 CI+MBPT method

The CI+MBPT method is employed for calculating atomic structure under the framework of relativity, which fully combines the advantages of configuration-interaction (CI) method and many-body perturbation theory (MBPT) method [24,25]. The CI method considers the correlation effect between valence electrons (VV) under the approximation of "frozen" core, while the MBPT method captures both the correlation effect between core electrons and the correlation effect between core and valence electrons [24,25]. The effectiveness of CI+MBPT has been extensively validated, with many studies demonstrating excellent agreement between calculated results and high-precision experimental measurements [24-32].

The wave function of the valence electronic states in the CI method $\Psi(\gamma PJ)$ consists of a linear combination of a set of configured wave functions with a specific angular momentum $J$ and a parity $P$. It can be expressed as follows [24,25],

$$\Psi(\gamma PJ) = \sum_{i=1}^{M} c_i \Phi(\gamma_i PJ), \quad (1).$$

Where, $\Phi(\gamma_i PJ)$ represents a configuration wave function in an M-dimensional CI model space, $c_i$ denotes the linear combination coefficient, and $\gamma_i$ represents the unspecified remaining quantum number. The CI model space encompasses wave functions constructed from the single (S) and double (D) excitation of valence electrons from the selected reference group to a certain imaginary orbit. The configuration wave function is defined by the Slater determinant given by the single-electron Dirac orbital. The single-electron Dirac orbital can be expressed as [24,25],

$$\phi_{nkm}(r,\theta,\varphi) = \frac{1}{r} \begin{pmatrix} P_{nk}(r) \cdot \Omega_{km}(\theta,\varphi) \\ iQ_{nk}(r) \cdot \Omega_{-km}(\theta,\varphi) \end{pmatrix}, \quad (2).$$

Where $P_{nk}(r)$ and $Q_{nk}(r)$ are the radial wave functions, $\Omega_{km}(\theta,\varphi)$ represents the angular wave function, $n$ is the principal quantum number, $k = \pm(j + 1/2)$ is the relativistic quantum number, and $m$ is the magnetic quantum number corresponding to the total angular momentum of a single electron. In this work, the reference configurations for the even parity states of Al II are $3s^2$, 3s4s and 3s3d, the reference configuration for the odd parity states is 3s3p, and the maximum orbitals of the single and double excitations are 29s, 29p, 29d and 29f. The single electron orbitals include 1s-29s, 2p-29p, 3d-29d and 4f-29f , where the real orbitals (n < 2) and 3s, 4s, 3p, 4p, 4p, 3d, 4d and 4f orbitals are obtained by solving the Dirac-Hartree-Fock (DHF) equation under the $V^{N-2}$ approximation, and the remaining imaginary orbitals are automatically generated by the recursive program [25]. In the trial calculation process, we performed a strict convergence test on the radial integration range ($R_{max}$ parameter), and finally selected the radial integration range as $R_{max}$ = 250 a.u..

The wave function of the valence electron state $n$ is determined by solving for the energy eigen equation. The energy eigen equation is expressed as [24, 25],

$$H\Psi_n = E_n \Psi_n, \quad (3).$$

The CI Hamiltonian of valence electrons takes the form [24-32],

$$H_{CI} = \sum_{i=1}^{2}[c\alpha_i \cdot p_i + (\beta_i - 1)c^2 + V_{nuc}(r_i) + V^{N-2}(r_i)] + \frac{1}{2}\sum_{i,j=1}^{2} V_c(r_{i,j}), \quad (4).$$

where $c$ denotes the speed of light in a vacuum, $\alpha$ and $\beta$ are Dirac matrices, $p$ represents the momentum operator, $V_{nuc}(r)$ is the nuclear potential, $V^{N-2}(r)$ describes the DHF potential from the core electrons, and $V_c(r_{1,2})$ denotes the inter-electron Coulomb potential at a distance of $r_{1,2}$. The CI+MBPT method extends this framework by incorporating all valence-core (VC) and core-core (CC) correlations into the effective Hamiltonian $H_{eff}$ through the many-body perturbation theory method. The effective Hamiltonian $H_{eff}$ can be expressed as follows [24, 25]:

$$H_{eff} = H_{CI} + \Sigma_1 + \Sigma_2, \quad (5).$$

Where, $\Sigma_1$ represents the single-electron correlation operator, describing the correlations between a single valence electron and the core; $\Sigma_2$ denotes the two-electron correlation operator, which describes the shielded Coulomb potential of the core for valence electrons. These correlation operators, treated to second order in MBPT, capture all valence-core and core-core correlations in the effective Hamiltonian. Solving the energy eigenvalue equation with $H_{eff}$ thus accounts for electron correlations throughout the atomic system. The CI+MBPT package used in this work was developed by M. G. Kozlov et al., and the specific program operation process can be referred to reference [25]. The quantum electrodynamic effects of interactions between electrons are described by the Breit interaction, with relevant functionality options embedded in the package [25].

**2.2 Transition probabilities and oscillator strengths**

We present general expressions for multipole transition properties in atomic systems, including transition rates, oscillator strengths, and line strengths. For transitions between an initial state i and final state k, the transition rates (in s⁻¹) for electric dipole (E1), electric quadrupole (E2), magnetic dipole (M1), and magnetic quadrupole (M2) transitions take the following forms [26, 33, 34],

$$A_{ki}^{E1} = \frac{2.02613 \times 10^{18}}{\lambda^3} \cdot \frac{S_{ik}^{E1}}{2J_i + 1} \, s^{-1}, \quad (6).$$

$$A_{ki}^{E2} = \frac{1.1199 \times 10^{18}}{\lambda^5} \cdot \frac{S_{ik}^{E2}}{2J_i + 1} \, s^{-1}, \quad (7).$$

$$A_{ki}^{M1} = \frac{2.69735 \times 10^{13}}{\lambda^3} \cdot \frac{S_{ik}^{M1}}{2J_i + 1} \; s^{-1}, \tag{8}$$

and

$$A_{ki}^{M2} = \frac{1.49097 \times 10^{13}}{\lambda^5} \cdot \frac{S_{ik}^{M2}}{2J_i + 1} \; s^{-1}, \tag{9}$$

The dimensionless oscillator strengths $f_{ki}$ are related to the transition rates $A_{ki}$ through $f_{ik} = 1.4992 \times 10^{-16} \times \frac{2J_k+1}{2J_i+1} A_{ki}\lambda^2$. We can derive the corresponding E1, E2, and M1 oscillator strengths as follows[33, 34],

$$f_{ik}^{E1} = \frac{303.756}{\lambda(2J_k + 1)} S_{ik}^{E1}, \tag{10}$$

$$f_{ik}^{E2} = \frac{167.90}{\lambda^3(2J_k + 1)} S_{ik}^{E2}, \tag{11}$$

$$f_{ik}^{M1} = \frac{0.00404386}{\lambda(2J_k + 1)} S_{ik}^{M1}, \tag{12}$$

and

$$f_{ik}^{M2} = \frac{0.00223526}{\lambda^3(2J_k + 1)} S_{ik}^{M2}, \tag{13}$$

The numerical factors above incorporate unit conversions and fundamental constants [26, 33, 34]. In these equations, $\lambda$ is the wavelength in vacuum of the transition measured in angstroms (Å), $S_{ik}^O = |\langle i||O||k\rangle|^2$ represents the line strength due to the corresponding transition type, and $\langle i||O||k\rangle$ represents the reduced transition matrix elements of corresponding transition.

## 3. Results and Discussion

### 3.1 Energies

We present energies of the ground state and 102 lowest excitation states (57 even states and 45 odd states) of Al II calculated using both pure CI method and CI+MBPT method. Results for even-parity and odd-parity states are presented in Tables 1 and 2, respectively, along with the effects of the Breit interaction. We compare our results with available experimental measurements, previous theoretical calculations [2, 15-20, 22, 34], and reference values from the NIST atomic spectra database [23].

Figure 1 compares our CI+MBPT calculated energies with NIST reference values, showing the relative differences as a function of state energy. Our CI+MBPT results demonstrate excellent agreement with both experimental measurements and previous theoretical calculations, with differences typically below 3%. Compared to NIST reference values, our calculations show deviations less than 0.2% for most states, with absolute differences generally below 100 a.u.. Larger deviations, exceeding 1% in four cases, occur primarily for states with high principal quantum numbers, which are known to present computational challenges due to their complex configuration structure [2, 8, 15-20, 23, 34, 35].

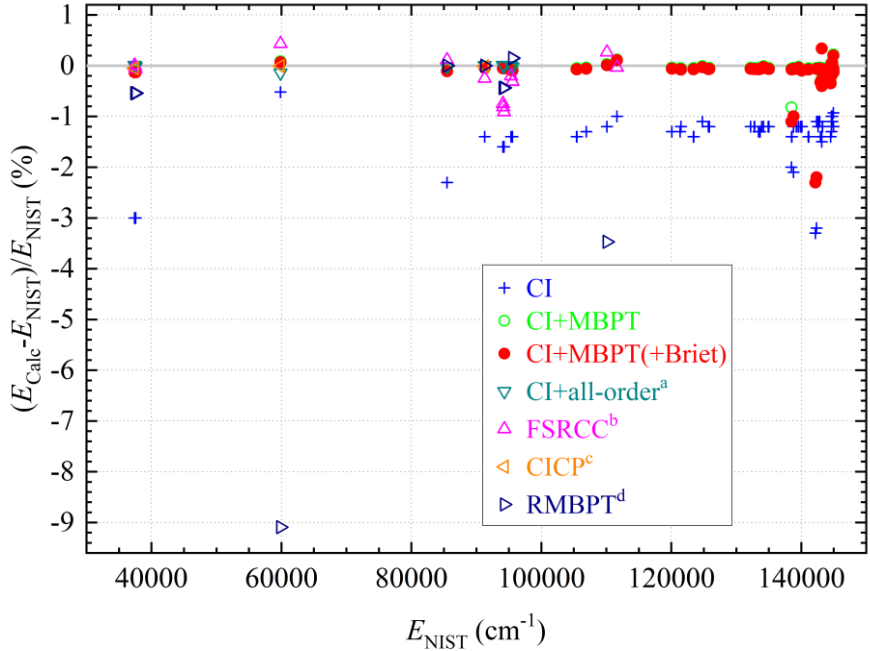

Figure 1. Comparison of energy level calculations between different methods. The horizontal axis shows NIST database values (in cm$^{-1}$), and the vertical axis shows the relative differences between calculated values (from CI+MBPT and other theoretical methods) and NIST reference data (in percentage). References: [a] Safronova *et al*. (2011) [2]; [b] Kumar *et al*. (2021) [19]; [c] Mitroy *et al*. (2009) [17]; [d] Johnson *et al*. (1997) [35].

The incorporation of many-body perturbation theory significantly improves the accuracy of our results, showing improvements of 1-2 orders of magnitude compared to pure CI calculations. The Breit interaction contributions are found to be small, with relative effects below 0.01%, indicating that the valence electron correlation model effectively describes the atomic structure of Al II. Our calculations successfully capture both valence and core electron correlations, producing accurate results even for challenging high-lying states. This systematical set of energy level calculations, including states with high principal quantum numbers and orbital angular momenta, provides valuable reference data for future experimental spectroscopic studies of Al II.

### 3.2 Reduced transition matrix elements

We calculate the RMEs of relevant transition using CI + MBPT method. In the calculation of the reduced matrix elements, we have also taken the Breit interaction into account. The RMEs of E1, E2, M1, and M2 transitions arising from 5 lowest energy states are presented in Table 3-6 (E1 in Table 3, E2 in Table 4, M1 in Table 5, and M2 in Table 6). The available theoretical and experimental values from the Refs. [2, 8, 15-20, 22, 34, 35] are also included for comparison. For the E1 RMEs, the length gauge and velocity gauge are both presented, as their agreement provides a valuable metric for assessing calculation error. Previous studies have established that the length gauge typically yields more accurate results for electric dipole transitions [2, 19].

Our calculated RMEs show good agreement with existing theoretical and experiment values [2, 8, 15-20, 22, 34, 35]. For stronger transitions, such as the $3s^2$ $^1S_0$-$3s3p$ $^1P_1$ transition, the difference values between our results and the other reference values are less than 10%, and for the weaker transitions, the relative difference values are less than 40%. According to the E1 transitions, most of the difference values between the length gauge and velocity gauge are less than 1% for stronger transitions, 10% for weaker transitions. As far as we know [2, 8, 15-20, 22, 34, 35], many of our calculations presented the first theoretical predictions for these transitions. The large magnitudes of these previously unreported RMEs, such as, $3s3p$ $^3P_0$-$3s3p$ $^3P_2$ (M2), $3s3p$ $^3P_0$-$3s4p$ $^3P_2$ (M2), suggest significant transition strengths, warranting future experimental investigation.

### 3.3 Transition probabilities and oscillator strengths

Using equations (6)-(13) and wavelengths from the NIST database, we calculate transition probabilities and oscillator strengths for the transitions presented in Table 3-6. Comparison with NIST reference data demonstrates excellent agreement, with most calculated values differing by less than 10% from NIST values. For more than ten transitions, the agreement is particularly strong, showing differences below 1%. Transitions that differ significantly from the data in the NIST-related databases are mostly weak transitions or have relatively high upper states energies. Equations (6) - (13) reveal that when only considering the change in the reduced matrix elements, the error in the transition probability is twice the error in the reduced matrix elements. An interesting case emerges for transitions such as $3s^2$ $^1S_0$-$3s3p$ $^3P_1$ (E1), where we observe substantial differences in reduced matrix elements compared to previous calculations, yet the resulting transition probabilities show remarkably small deviations from NIST database values. This consistency in transition probabilities provides strong validation of our computational approach and offers valuable insights for uncertainty estimation in Al II precision measurements, particularly for quantum optical clock experiments. Furthermore, our work extends beyond existing NIST data by providing predictions for previously unreported transitions, including $3s3p$ $^3P_0$-$3s3p$ $^3P_1$ (M1), $3s^2$ $^1S_0$-$3s3d$ $^1D_2$ (E2), et. These new results establish important theoretical benchmarks for future experimental and theoretical investigations of Al II.

## 4. Conclusions

We have conducted comprehensive relativistic calculations of atomic properties for Al II ion using the configuration interaction plus many-body perturbation theory (CI+MBPT) method. We present accurate energy levels for the 103 lowest-lying states of Al II, and comprehensive multipole transition data, including reduced matrix elements, transition probabilities, and oscillator strengths, for transitions originating from the five most important low-lying configurations ($3s^2$ $^1S_0$, $3s3p$ $^3P_0$, $3s3p$ $^3P_1$, $3s3p$ $^3P_2$, and $3s3p$ $^1P_1$) among these levels. Our calculated energy levels demonstrate excellent agreement with values taken from the NIST atomic spectra database, showing deviations below 0.2% for most states. The computed transition properties show good agreement with available theoretical and experimental values, with differences typically below 10% and achieving better than 1% agreement for strong transitions. Notably, our results predict transition properties for numerous previously unexplored transitions, substantially expanding the available atomic data for Al II. These high-accuracy predictions establish important benchmarks for future Al II precision measurement and astrophysical spectroscopy applications.

# Acknowledgments

We would like to thank Prof. M. G. Kozlov of Petersburg Nuclear Physics Institute, Dr. Y. M. Yu of Institute of Physics, Chinese Academy of Sciences, and Dr. C.-B. Li of Innovation Academy for Precision Measurement Science and Technology, Chinese Academy of Sciences for the helpful assistance on the use of CI+MBPT package. We also thank Dr. J. S. Liu and Prof. Q. T. Meng of Shandong Normal University and Prof. C. Z. Dong from Northwest Normal University for the helpful discussion. This work acknowledges financial support from Hubei Provincial Natural Science Foundation (HPNSF) Excellence Group Program (Grant No.JCKXXM202500020), the Science and Technology Development Fund (FDCT), Macao SAR (Grant No. 0024/2024/RIB1), Shanghai Municipal Science and Technology Major Project (Grant No. 2019SHZDZX01), and the National Natural Science Foundation of China (Grant No. 12404421).

# Declaration of generative AI and AI-assisted technologies in the writing process

During the preparation of this work the authors used Claude 3.5 Sonnet in order to proofread and improve language. After using this tool, the authors reviewed and edited the content as needed and take full responsibility for the content of the published article.

# Explanation of Tables

In Tables 1 and 2, we present the energies of the ground state and 102 lowest-energy excited states for Al II, with Table 1 containing even states and Table 2 containing odd states. Tables 3-6 provide transition wavelengths, reduced matrix elements, transition probabilities, and oscillator strengths for approximately 400 electric dipole (E1), magnetic dipole (M1), electric quadrupole (E2), and magnetic quadrupole (M2) transitions arising from these levels. Comparisons with values from existing literature are included in all tables.

**Table 1. Energies of 58 lowest even states of Al II (cm$^{-1}$).**

**Table 2. Energies of 45 lowest odd states of Al II (cm$^{-1}$).**

| | |
|---|---|
| First column | The configuration of the states for Al II. |
| Second column | The terms of the states for Al II. |
| Third and fourth columns | Energies in cm$^{-1}$ getting by CI and difference with NISTDATA in %. |
| Fifth and sixth columns | Energies in cm$^{-1}$ getting by CI+MBPT and difference with NISTDATA in %. |
| Seventh and eighth columns | Energies in cm$^{-1}$ getting by CI+MBPT considering the Breit interactions and difference with NISTDATA in %. |
| Ninth column | Energies of NISTDATA in cm$^{-1}$. |
| Tenth column | Energies of other literature in cm$^{-1}$. |
| Config. | Configuration. |
| Ref. | Reference. |
| Δ | The difference of NISTDATA. |

**Table 3. The transition amplitudes of the E1 transitions arising from 5 lowest energy states.**

| | |
|---|---|
| First and Second column | Transition levels, the upper (initial) and the lower (final) levels. |
| Third column | Transition wavelength $\lambda$, in nm. |
| Fourth, fifth and sixth columns | Our values of RMEs in length (L) and velocity (V) forms of the E1 operator in a.u. and other literature values. |
| Seventh to Tenth columns | Our values of transition rates in s$^{-1}$ in length (L) and velocity (V) forms of the E1 operator and the difference with NISTDATA in % |
| Eleventh column | Transition rates of NISTDATA in s$^{-1}$. |
| Twelfth column | Our values of oscillator strengths in length (L) and velocity (V) forms of the E1 operator. |
| L | Length form of the electric-dipole operator. |
| V | Velocity form of the electric-dipole operator. |
| Diff. | The difference of NISTDATA. |

**Table 4**. **The transition amplitudes of the M1 transitions arising from 5 lowest energy states.**

**Table 5. The transition amplitudes of the E2 transitions arising from 5 lowest energy states.**

**Table 6. The transition amplitudes of the M2 transitions arising from 5 lowest energy states.**

| | |
|---|---|
| First and Second column | Transition levels, the upper (initial) and the lower (final) levels. |
| Third column | Transition wavelength $\lambda$, in nm. |
| Fourth and fifth columns | Our values of RMEs in a.u. and other literature values. |
| Sixth and seventh columns | Our values of transition rates in s$^{-1}$ and the difference with NISTDATA in %. |
| Eighth column | Transition rates of NISTDATA in s$^{-1}$. |
| Ninth column | Our values of oscillator strengths. |

| Config. | Term | CI | | CI+MBPT | | +Briet | | NISTDATA | Ref. |
|---|---|---|---|---|---|---|---|---|---|
| | | Δ | % | Δ | % | Δ | % | | |
| $3s^2$ | $^1S_0$ | 376616.91 | -1.2 | 381042.87 | -0.070 | 380973.00 | -0.088 | 381308.21 | 381332 [2] |
| | | | | | | | | | 379582 [19] |
| | | | | | | | | | 381287 [17] |
| | | | | | | | | | 382024 [35] |
| $3p^2$ | $^1D_2$ | 83515.32 | -2.3 | 85404.27 | -0.090 | 85384.30 | -0.11 | 85481.35 | 85465 [2] |
| | | | | | | | | | 85578 [19] |
| | | | | | | | | | 85678 [35] |
| $3s4s$ | $^3S_1$ | 90007.76 | -1.4 | 91253.54 | -0.023 | 91233.34 | -0.045 | 91274.50 | 91279 [2] |
| | | | | | | | | | 91043 [19] |
| | | | | | | | | | 91262 [17] |
| | | | | | | | | | 91274 [35] |
| $3p^2$ | $^3P_0$ | 92597.07 | -1.6 | 94041.20 | -0.047 | 94034.25 | -0.054 | 94084.96 | 94092 [2] |
| | | | | | | | | | 93379 [19] |
| | | | | | | | | | 93672 [35] |
| $3p^2$ | $^3P_1$ | 92659.50 | -1.6 | 94107.24 | -0.042 | 94097.22 | -0.053 | 94147.46 | 94151 [2] |
| | | | | | | | | | 93380 [19] |
| | | | | | | | | | 93735 [35] |
| $3p^2$ | $^3P_2$ | 92781.51 | -1.6 | 94236.13 | -0.034 | 94219.44 | -0.052 | 94268.68 | 94265 [2] |
| | | | | | | | | | 93409 [19] |
| | | | | | | | | | 93857 [35] |
| $3s4s$ | $^1S_0$ | 94006.78 | -1.4 | 95289.06 | -0.065 | 95268.00 | -0.087 | 95350.60 | 95354 [2] |
| | | | | | | | | | 95156 [19] |
| $3s3d$ | $^3D_3$ | 94169.47 | -1.4 | 95490.21 | -0.062 | 95460.82 | -0.093 | 95549.42 | 95524 [2] |
| | | | | | | | | | 95253 [19] |
| | | | | | | | | | 95690 [35] |
| $3s3d$ | $^3D_2$ | 94170.17 | -1.4 | 95490.47 | -0.062 | 95461.61 | -0.093 | 95550.51 | 95527 [2] |
| | | | | | | | | | 95252 [19] |
| | | | | | | | | | 95697 [35] |
| $3s3d$ | $^3D_1$ | 94170.62 | -1.4 | 95490.62 | -0.062 | 95462.35 | -0.093 | 95551.44 | 95527 [2] |
| | | | | | | | | | 95248 [19] |
| | | | | | | | | | 95695 [35] |
| $3s3d$ | $^1D_2$ | 108816.30 | -1.2 | 110121.70 | 0.030 | 110099.29 | 0.0086 | 110089.83 | 110382 [19] |
| | | | | | | | | | 106270 [35] |
| $3p^2$ | $^1S_0$ | 110488.16 | -1.0 | 111779.38 | 0.12 | 111761.18 | 0.11 | 111637.33 | 111598 [19] |
| $3s5s$ | $^3S_1$ | 118590.46 | -1.3 | 120046.91 | -0.038 | 120022.48 | -0.059 | 120092.92 | |
| $3s5s$ | $^1S_0$ | 119827.45 | -1.3 | 121305.27 | -0.051 | 121280.64 | -0.071 | 121366.73 | |
| $3s4d$ | $^3D_3$ | 119970.30 | -1.2 | 121420.01 | -0.052 | 121392.32 | -0.077 | 121483.50 | |
| $3s4d$ | $^3D_2$ | 119970.54 | -1.2 | 121421.41 | -0.051 | 121393.94 | -0.074 | 121483.92 | |
| $3s4d$ | $^3D_1$ | 119970.71 | -1.2 | 121422.34 | -0.051 | 121395.10 | -0.073 | 121484.25 | |
| $3s4d$ | $^1D_1$ | 123376.57 | -1.1 | 124783.82 | -0.0083 | 124756.80 | -0.030 | 124794.13 | |
| $3s6s$ | $^3S_1$ | 130638.67 | -1.2 | 132160.28 | -0.041 | 132134.49 | -0.061 | 132215.52 | |
| $3s6s$ | $^1S_0$ | 131185.11 | -1.2 | 132715.46 | -0.048 | 132689.58 | -0.067 | 132778.63 | |

| | | | | | | | | |
|---|---|---|---|---|---|---|---|---|
| 3s5d | $^3D_2$ | 131247.03 | -1.2 | 132760.80 | -0.047 | 132733.58 | -0.067 | 132822.80 |
| 3s5d | $^3D_3$ | 131247.09 | -1.2 | 132760.34 | -0.047 | 132732.99 | -0.068 | 132822.89 |
| 3s5d | $^3D_1$ | 131246.99 | -1.2 | 132761.11 | -0.047 | 132734.00 | -0.067 | 132822.95 |
| 3s5d | $^1D_1$ | 132359.24 | -1.2 | 133873.46 | -0.032 | 133845.98 | -0.053 | 133916.37 |
| 3s5g | $^3G_4$ | 132585.60 | -1.2 | 134172.32 | -0.0085 | 134145.22 | -0.029 | 134183.7 |
| 3s5g | $^3G_5$ | 132585.82 | -1.2 | 134172.53 | -0.0083 | 134145.39 | -0.029 | 134183.7 |
| 3s5g | $^3G_3$ | 132585.50 | -1.2 | 134172.23 | -0.0085 | 134145.14 | -0.029 | 134183.7 |
| 3s5g | $^1G_4$ | 132585.95 | -1.2 | 134172.66 | -0.0082 | 134145.53 | -0.029 | 134183.7 |
| 3s7s | $^3S_1$ | 135798.86 | -2.0 | 137364.03 | -0.82 | 137337.07 | -1.1 | 138499.89 |
| 3s7s | $^1S_0$ | 135843.45 | -2.1 | 137410.35 | -1.0 | 137383.35 | -1.0 | 138800.60 |
| 3s6d | $^3D_1$ | 137209.94 | -1.2 | 138753.58 | -0.044 | 138726.50 | -0.064 | 138814.87 |
| 3s6d | $^3D_2$ | 137210.06 | -1.2 | 138753.51 | -0.044 | 138726.36 | -0.064 | 138814.87 |
| 3s6d | $^3D_3$ | 137210.25 | -1.2 | 138753.39 | -0.044 | 138726.16 | -0.064 | 138815.12 |
| 3s6d | $^1D_1$ | 137681.84 | -1.2 | 139235.78 | -0.038 | 139208.42 | -0.058 | 139289.15 |
| 3s6g | $^3G_3$ | 137973.63 | -1.2 | 139559.33 | -0.023 | 139532.24 | -0.042 | 139591.1 |
| 3s6g | $^3G_5$ | 137973.88 | -1.2 | 139559.57 | -0.023 | 139532.45 | -0.042 | 139591.1 |
| 3s6g | $^3G_4$ | 137973.71 | -1.2 | 139559.41 | -0.023 | 139532.31 | -0.042 | 139591.1 |
| 3s6g | $^1G_4$ | 137973.99 | -1.2 | 139559.68 | -0.023 | 139532.56 | -0.042 | 139591.1 |
| 3s8s | $^3S_1$ | 137433.80 | -3.3 | 138974.85 | -2.3 | 138948.52 | -2.3 | 142182.94 |
| 3s8s | $^1S_0$ | 137740.27 | -3.2 | 139285.66 | -2.2 | 139259.32 | -2.2 | 142363.06 |
| 3s7d | $^3D_1$ | 140741.49 | -1.1 | 142300.66 | -0.046 | 142273.58 | -0.065 | 142365.54 |
| 3s7d | $^3D_2$ | 140742.68 | -1.1 | 142301.67 | -0.045 | 142274.55 | -0.064 | 142365.69 |
| 3s7d | $^3D_3$ | 140744.47 | -1.1 | 142303.19 | -0.044 | 142276.01 | -0.063 | 142365.98 |
| 3s7d | $^1D_2$ | 140979.43 | -1.1 | 142549.38 | -0.042 | 142522.11 | -0.061 | 142609.27 |
| 3s7g | $^3G_5$ | 141213.70 | -1.1 | 142798.48 | -0.037 | 142771.37 | -0.056 | 142851.6 |
| 3s7g | $^3G_3$ | 141209.16 | -1.1 | 142793.91 | -0.040 | 142766.82 | -0.059 | 142851.6 |
| 3s7g | $^3G_4$ | 141209.37 | -1.1 | 142794.11 | -0.040 | 142767.00 | -0.059 | 142851.6 |
| 3s7g | $^1G_3$ | 141213.62 | -1.1 | 142798.41 | -0.037 | 142771.32 | -0.056 | 142851.6 |
| 3s9g | $^3S_1$ | 142509.36 | -1.4 | 144046.14 | -0.33 | 144019.89 | -0.35 | 144527.35 |
| 3s8d | $^3D_1$ | 143117.86 | -1.1 | 144685.91 | 0.030 | 144658.84 | 0.012 | 144642.0 |
| 3s8d | $^3D_2$ | 143130.64 | -1.0 | 144698.44 | 0.039 | 144671.33 | 0.020 | 144642.18 |
| 3s8d | $^3D_3$ | 143154.37 | -1.0 | 144721.31 | 0.055 | 144694.15 | 0.036 | 144642.45 |
| 3s9s | $^1S_0$ | 142834.76 | -1.3 | 144376.75 | -0.18 | 144350.48 | -0.20 | 144644.14 |
| 3s8d | $^1D_2$ | 143306.26 | -1.0 | 144885.45 | 0.071 | 144858.21 | 0.052 | 144782.23 |
| 3s8g | $^3G_5$ | 143291.05 | -1.2 | 144874.88 | -0.064 | 144847.78 | -0.082 | 144967.1 |
| 3s8g | $^3G_4$ | 143215.19 | -1.2 | 144798.84 | -0.12 | 144771.73 | -0.13 | 144967.1 |
| 3s8g | $^3G_3$ | 143215.03 | -1.2 | 144798.68 | -0.12 | 144771.60 | -0.13 | 144967.1 |
| 3s8g | $^1G_4$ | 143290.98 | -1.2 | 144874.82 | -0.063 | 144847.73 | -0.086 | 144967.1 |

**Table 1.** Energies of 58 lowest even states of Al II (cm$^{-1}$).

| Config. | Level | CI | | CI+MBPT | | +Briet | | NISTDATA | Ref. |
|---|---|---|---|---|---|---|---|---|---|
| | | Δ | % | Δ | % | Δ | % | | |
| 3s3p | $^3P_0$ | 36255.56 | -3.0 | 37342.30 | -0.13 | 37344.07 | -0.13 | 37393.03 | 37396 [2] |
| | | | | | | | | | 37395 [19] |
| | | | | | | | | | 37374 [17] |
| | | | | | | | | | 37191 [35] |
| 3s3p | $^3P_1$ | 36317.75 | -3.0 | 37407.30 | -0.12 | 37405.28 | -0.13 | 37453.91 | 37457 [2] |
| | | | | | | | | | 37452 [19] |
| | | | | | | | | | 37251 [35] |
| 3s3p | $^3P_2$ | 36442.90 | -3.0 | 37538.57 | -0.10 | 37530.14 | -0.13 | 37577.79 | 37572 [2] |
| | | | | | | | | | 37555 [19] |
| | | | | | | | | | 37374 [35] |
| 3s3p | $^1P_1$ | 59538.18 | -0.52 | 59904.79 | 0.088 | 59893.43 | 0.069 | 59852.02 | 59768 [2] |
| | | | | | | | | | 60111 [19] |
| | | | | | | | | | 59848 [17] |
| | | | | | | | | | 54410 [35] |
| 3s4p | $^3P_0$ | 103960.74 | -1.4 | 105372.09 | -0.053 | 105352.32 | -0.071 | 105427.52 | |
| 3s4p | $^3P_1$ | 103975.18 | -1.4 | 105387.13 | -0.052 | 105366.49 | -0.071 | 105441.50 | |
| 3s4p | $^3P_2$ | 104004.94 | -1.4 | 105418.40 | -0.050 | 105396.26 | -0.071 | 105470.93 | |
| 3s4p | $^1P_1$ | 105573.59 | -1.3 | 106879.60 | -0.038 | 106858.31 | -0.058 | 106920.56 | |
| 3s4f | $^3F_2$ | 121692.99 | -1.4 | 123358.35 | -0.049 | 123330.94 | -0.071 | 123418.48 | |
| 3s4f | $^3F_3$ | 121695.55 | -1.4 | 123360.63 | -0.048 | 123333.03 | -0.071 | 123420.45 | |
| 3s4f | $^3F_4$ | 121699.08 | -1.4 | 123363.81 | -0.048 | 123335.99 | -0.071 | 123423.36 | |
| 3s4f | $^1F_3$ | 121774.86 | -1.4 | 123413.08 | -0.047 | 123385.49 | -0.069 | 123470.5 | |
| 3s5p | $^3P_0$ | 124135.95 | -1.2 | 125641.43 | -0.049 | 125617.47 | -0.068 | 125703.14 | |
| 3s5p | $^3P_1$ | 124141.90 | -1.2 | 125647.85 | -0.049 | 125623.56 | -0.068 | 125708.83 | |
| 3s5p | $^3P_2$ | 124155.03 | -1.2 | 125662.46 | -0.047 | 125637.49 | -0.067 | 125721.70 | |
| 3s5p | $^1P_1$ | 124322.49 | -1.2 | 125814.33 | -0.043 | 125789.78 | -0.063 | 125869.02 | |
| 3s5f | $^3F_2$ | 131656.54 | -1.3 | 133371.20 | -0.050 | 133343.70 | -0.070 | 133437.71 | |
| 3s5f | $^3F_3$ | 131663.70 | -1.3 | 133376.91 | -0.050 | 133349.06 | -0.071 | 133443.08 | |
| 3s5f | $^3F_4$ | 131673.18 | -1.3 | 133384.50 | -0.049 | 133356.23 | -0.070 | 133450.07 | |
| 3s5f | $^1F_3$ | 132002.28 | -1.3 | 133621.30 | -0.045 | 133593.87 | -0.066 | 133681.78 | |
| 3s6p | $^1P_1$ | 133307.71 | -1.2 | 134861.37 | -0.043 | 134835.67 | -0.062 | 134919.40 | |
| 3s6p | $^3P_0$ | 133405.88 | -1.2 | 134950.18 | -0.046 | 134924.73 | -0.065 | 135012.29 | |
| 3s6p | $^3P_1$ | 133409.43 | -1.2 | 134953.89 | -0.046 | 134928.22 | -0.065 | 135015.70 | |
| 3s6p | $^3P_2$ | 133416.13 | -1.2 | 134960.79 | -0.045 | 134934.81 | -0.065 | 135022.13 | |
| 3s6f | $^3F_2$ | 136544.63 | -1.4 | 138443.93 | -0.056 | 138415.62 | -0.076 | 138521.4 | |
| 3s6f | $^3F_3$ | 136567.69 | -1.4 | 138462.95 | -0.055 | 138433.57 | -0.076 | 138538.9 | |
| 3s6f | $^3F_4$ | 136597.82 | -1.4 | 138487.68 | -0.053 | 138457.02 | -0.076 | 138561.8 | |
| 3s6f | $^1F_3$ | 137576.04 | -1.2 | 139184.26 | -0.044 | 139156.93 | -0.063 | 139245.34 | |
| 3s7p | $^1P_1$ | 138244.62 | -1.2 | 139828.16 | -0.065 | 139801.86 | -0.084 | 139918.98 | |
| 3s7p | $^3P_0$ | 138416.51 | -1.2 | 139980.55 | -0.078 | 139954.38 | -0.097 | 140090.0 | |
| 3s7p | $^3P_1$ | 138422.45 | -1.2 | 139986.66 | -0.075 | 139960.37 | -0.094 | 140091.9 | |
| 3s7p | $^3P_2$ | 138433.86 | -1.2 | 139998.24 | -0.070 | 139971.76 | -0.088 | 140095.7 | |
| Config. | Level | CI | | CI+MBPT | | +Briet | | NISTDATA | Ref. |
| | | Δ | % | Δ | % | Δ | % | | |

| | | | | | | | | |
|---|---|---|---|---|---|---|---|---|
| 3p3d | $^3F_2$ | 139141.69 | -1.4 | 141008.87 | -0.054 | 140980.05 | -0.074 | 141084.89 |
| 3p3d | $^3F_3$ | 139163.31 | -1.4 | 141035.56 | -0.053 | 141005.15 | -0.074 | 141110.06 |
| 3p3d | $^3F_4$ | 139191.41 | -1.4 | 141070.51 | -0.051 | 141038.14 | -0.074 | 141143.05 |
| 3s7f | $^1F_3$ | 140934.75 | -1.2 | 142536.12 | -0.048 | 142508.85 | -0.067 | 142604.05 |
| 3s8p | $^1P_1$ | 140927.37 | -1.4 | 142519.32 | -0.31 | 142492.70 | -0.33 | 142961.20 |
| 3s8p | $^3P_0$ | 141049.25 | -1.5 | 142619.89 | -0.38 | 142593.40 | -0.40 | 143165.4 |
| 3s8p | $^3P_1$ | 141126.76 | -1.4 | 142711.45 | -0.32 | 142684.89 | 0.34 | 143166.76 |
| 3s8p | $^3P_2$ | 141207.88 | -1.4 | 142791.99 | -0.26 | 142765.28 | -0.28 | 143170.04 |
| 3s7f | $^3F_2$ | 141537.90 | -1.2 | 143188.01 | -0.054 | 143160.17 | -0.074 | 143265.83 |
| 3s7f | $^3F_3$ | 141546.75 | -1.2 | 143199.08 | -0.051 | 143170.82 | -0.071 | 143272.86 |
| 3s7f | $^3F_4$ | 141559.19 | -1.2 | 143215.45 | -0.048 | 143186.50 | -0.068 | 143283.75 |
| 3s7f | $^1F_3$ | 143179.40 | -1.1 | 144780.03 | -0.0031 | 144752.80 | -0.022 | 144784.45 |
| 3s9p | $^1P_1$ | 143590.88 | -0.93 | 145259.50 | 0.22 | 145232.99 | 0.20 | 144941.10 |

**Table 2.** Energies of 45 lowest odd states of Al II (cm$^{-1}$).

| Transition | | λ (nm) | RMEs (a. u.) | | | $A_{ki}$ ($s^{-1}$) | | | | | $f_{ik}$ | |
|---|---|---|---|---|---|---|---|---|---|---|---|---|
| Initial | Final | | L | V | Ref. | L | Diff (%) | V | Diff (%) | NISTDATA | L | V |
| $3s^2\ ^1S_0$ | $3s3p\ ^3P_1$ | 266.9948211 | 9.8467E-3 | 0.010517 | 0.01513 [19] | 3.4405E+3 | 4.9 | 3.9248E+3 | 20 | 3.28E+3 | 1.1031E-5 | 1.2584E-5 |
| $3s^2\ ^1S_0$ | $3s3p\ ^1P_1$ | 167.0787385 | 3.1150 | 3.1682 | 3.113 [2] | 1.4051E+9 | -0.35 | 1.4535E+9 | 3.1 | 1.41E+9 | 1.7641 | 1.8249 |
| | | | | | 2.840 [19] | | | | | | | |
| $3s^2\ ^1S_0$ | $3s4p\ ^3P_1$ | 94.8393184 | 2.2769E-3 | 1.7920E-3 | | 4.1046E+3 | | 2.5425E+3 | | | 1.6605E-6 | 1.0286E-6 |
| $3s^2\ ^1S_0$ | $3s4p\ ^1P_1$ | 93.5273814 | 0.045764 | 0.073706 | 0.045 [2] | 1.7289E+6 | | 4.4847E+6 | | | 6.8022E-4 | 1.7644E-3 |
| $3s^2\ ^1S_0$ | $3s5p\ ^3P_1$ | 79.5489067 | 5.3404E-3 | -4.2942E-3 | | 3.8264E+4 | | 2.4740E+4 | | | 1.0891E-5 | 7.0416E-6 |
| $3s^2\ ^1S_0$ | $3s5p\ ^1P_1$ | 79.4476671 | 0.066205 | 0.049113 | | 5.9031E+6 | | 3.2486E+6 | | | 1.6759E-3 | 9.2225E-4 |
| $3s^2\ ^1S_0$ | $3s6p\ ^1P_1$ | 74.1183254 | 0.070401 | 0.058554 | | 8.2210E+6 | | 5.6870E+6 | | | 2.0313E-3 | 1.4052E-3 |
| $3s^2\ ^1S_0$ | $3s6p\ ^3P_1$ | 74.0654605 | 2.2525E-3 | 1.7238E-3 | | 8.4339E+3 | | 4.9394E+3 | | | 2.0809E-6 | 1.2187E-6 |
| $3s^2\ ^1S_0$ | $3s7p\ ^1P_1$ | 71.469932 | 0.063805 | 0.055092 | | 7.5316E+6 | | 5.6150E+6 | | | 1.7303E-3 | 1.2900E-3 |
| $3s^2\ ^1S_0$ | $3s7p\ ^3P_1$ | 71.381714 | 2.1861E-3 | 1.8725E-3 | | 8.8741E+3 | | 6.5107E+3 | | | 2.0337E-6 | 1.4921E-6 |
| $3s^2\ ^1S_0$ | $3s8p\ ^1P_1$ | 69.9490491 | 0.050539 | 0.044846 | | 5.0403E+6 | | 3.9687E+6 | | | 1.1092E-3 | 8.7338E-4 |
| $3s^2\ ^1S_0$ | $3s8p\ ^3P_1$ | 69.8486156 | 0.024989 | 0.022049 | | 1.2376E+6 | | 9.6350E+5 | | | 2.7157E-4 | 2.1143E-4 |
| $3s^2\ ^1S_0$ | $3s9p\ ^1P_1$ | 68.9935428 | 0.078369 | 0.071839 | | 1.2630E+7 | | 1.0613E+7 | | | 2.7041E-3 | 2.2722E-3 |
| $3s3p\ ^3P_0$ | $3s4s\ ^3S_1$ | 185.5925608 | 0.89793 | 0.89255 | 0.900 [2] | 8.5182E+7 | 1.6 | 8.4165E+7 | 0.44 | 8.38E+7 | 0.13197 | 0.13039 |
| $3s3p\ ^3P_0$ | $3p^2\ ^3S_1$ | 176.1976994 | 1.8395 | 1.8788 | 1.836 [2] | 4.1778E+8 | 1.4 | 4.3582E+8 | 5.8 | 4.12E+8 | 0.58336 | 0.60855 |
| $3s3p\ ^3P_0$ | $3s3d\ ^3D_1$ | 171.9441779 | 2.2350 | 2.2626 | 2.236 [2] | 6.6365E+8 | 1.3 | 6.8014E+8 | 3.8 | 6.55E+8 | 0.88248 | 0.90441 |
| $3s3p\ ^3P_0$ | $3s5s\ ^3S_1$ | 120.9191451 | 0.26899 | 0.26610 | | 2.7640E+7 | 10 | 2.7049E+7 | 7.8 | 2.51E+7 | 0.018177 | 0.017788 |
| $3s3p\ ^3P_0$ | $3s4d\ ^3D_1$ | 118.9184792 | 0.44564 | 0.46120 | | 7.9756E+7 | -3.8 | 8.5423E+7 | 3.0 | 8.29E+7 | 0.050729 | 0.054333 |
| $3s3p\ ^3P_0$ | $3s6s\ ^3S_1$ | 105.460213 | 0.15053 | 0.14861 | | 1.3047E+7 | 11 | 1.2717E+7 | 7.8 | 1.18E+7 | 6.5267E-3 | 6.3613E-3 |
| $3s3p\ ^3P_0$ | $3s5d\ ^3D_1$ | 104.7889383 | 0.19214 | 0.20285 | | 2.1669E+7 | -7 | 2.4152E+7 | 3.7 | 2.33E+7 | 0.010702 | 0.011928 |
| $3s3p\ ^3P_0$ | $3s7s\ ^3S_1$ | 98.9052572 | 0.041434 | 0.042167 | | 1.1984E+6 | | 1.2412E+6 | | | 5.2727E-4 | 5.4609E-4 |
| $3s3p\ ^3P_0$ | $3s6d\ ^3D_1$ | 98.5980928 | 0.10968 | 0.11778 | | 8.4761E+6 | -8.5 | 9.7743E+6 | 5.6 | 9.26E+6 | 3.7062E-3 | 4.2738E-3 |
| $3s3p\ ^3P_0$ | $3s8s\ ^3S_1$ | 95.4290351 | 0.10273 | 0.10045 | | 8.2016E+6 | | 7.8416E+6 | | | 3.3593E-3 | 3.2119E-3 |

| Lower | Upper | λ (nm) | | | | | | | | | |
|---|---|---|---|---|---|---|---|---|---|---|---|
| 3s3p $^3P_0$ | 3s7d $^3D_1$ | 95.263036 | 0.072194 | 0.078691 | 4.0717E+6 | | 4.8375E+6 | | | 1.6619E-3 | 1.9745E-3 |
| 3s3p $^3P_0$ | 3s9s $^3S_1$ | 93.3407707 | 0.099132 | 0.098571 | 8.1613E+6 | | 8.0692E+6 | | | 3.1981E-3 | 3.1620E-3 |
| 3s3p $^3P_0$ | 3s8d $^3D_1$ | 93.2409887 | 0.057101 | 0.063005 | 2.7165E+6 | | 3.3073E+6 | | | 1.0622E-3 | 1.2932E-3 |
| 3s3p $^3P_1$ | $3p^2\ ^1D_2$ | 208.2143041 | 0.02085 | 0.021211 | 1.9515E+4 | | 2.0197E+4 | | | 2.1141E-5 | 2.1879E-5 |
| 3s3p $^3P_1$ | 3s4s $^1S_1$ | 185.8024967 | 1.5506 | 1.5411 | 2.5316E+8 | 1.7 | 2.5007E+8 | 0.43 | 2.49E+8 | 0.13103 | 0.12943 |
| 3s3p $^3P_1$ | $3p^2\ ^3P_0$ | 176.5815749 | 1.8376 | 1.8769 | 1.2426E+9 | 1.0 | 1.2963E+9 | 5.4 | 1.23E+9 | 0.19363 | 0.20200 |
| 3s3p $^3P_1$ | $3p^2\ ^3P_1$ | 176.3869082 | 1.5810 | 1.6150 | 3.0762E+8 | 1.2 | 3.2099E+8 | 5.6 | 3.04E+8 | 0.14349 | 0.14973 |
| 3s3p $^3P_1$ | $3p^2\ ^3P_2$ | 176.010569 | 2.0665 | 2.1103 | 3.1736E+8 | 1.4 | 3.3096E+8 | 5.7 | 3.13E+8 | 0.24567 | 0.25619 |
| 3s3p $^3P_1$ | 3s4s $^1S_0$ | 172.7214457 | 0.020601 | 0.021000 | 1.6688E+5 | | 1.7341E+5 | | | 2.4880E-5 | 2.5853E-5 |
| 3s3p $^3P_1$ | 3s3d $^3D_2$ | 172.1271124 | 3.3537 | 3.3951 | 8.9371E+8 | 1.3 | 9.1591E+8 | 3.8 | 8.82E+8 | 0.66163 | 0.67807 |
| 3s3p $^3P_1$ | 3s3d $^3D_1$ | 172.124357 | 1.9455 | 1.9695 | 5.0128E+8 | 2.0 | 5.1372E+8 | 4.2 | 4.93E+8 | 0.22266 | 0.22818 |
| 3s3p $^3P_1$ | 3s3d $^1D_2$ | 137.6729309 | 7.7504E-3 | 7.8463E-3 | 9.3282E+3 | | 9.5605E+3 | | | 4.4179E-6 | 4.5279E-6 |
| 3s3p $^3P_1$ | $3p^2\ ^1S_0$ | 134.8010107 | 9.8701E-3 | 0.010036 | 8.0580E+4 | | 8.3312E+4 | | | 7.3176E-6 | 7.5656E-6 |
| 3s3p $^3P_1$ | 3s5s $^3S_1$ | 121.008226 | 0.46678 | 0.46177 | 8.3047E+7 | 11 | 8.1274E+7 | 8.5 | 7.47E+7 | 0.018233 | 0.017842 |
| 3s3p $^3P_1$ | 3s5s $^1S_0$ | 119.1713018 | 2.0844E-4 | 2.6484E-4 | 52.013 | | 83.969 | | | 3.6915E-9 | 5.9595E-9 |
| 3s3p $^3P_1$ | 3s4d $^3D_2$ | 119.005103 | 0.66764 | 0.69108 | 1.0717E+8 | -4.3 | 1.1483E+8 | 2.5 | 1.12E+8 | 0.037926 | 0.040636 |
| 3s3p $^3P_1$ | 3s4d $^3D_1$ | 119.0046357 | 0.38542 | 0.39890 | 5.9528E+7 | -4.3 | 6.3765E+7 | 2.5 | 6.22E+7 | 0.012639 | 0.013539 |
| 3s3p $^3P_1$ | 3s4d $^1D_2$ | 114.4947883 | 7.2443E-3 | 7.2996E-3 | 1.4169E+4 | | 1.4386E+4 | | | 4.6411E-6 | 4.7123E-6 |
| 3s3p $^3P_1$ | 3s6s $^3S_1$ | 105.5279664 | 0.26111 | 0.25780 | 3.9182E+7 | 11 | 3.8195E+7 | 8.2 | 3.53E+7 | 6.5418E-3 | 6.3770E-3 |
| 3s3p $^3P_1$ | 3s6s $^1S_0$ | 104.9045829 | 2.7587E-4 | 2.8788E-4 | 133.56 | | 145.45 | | | 7.3457E-9 | 7.9992E-9 |
| 3s3p $^3P_1$ | 3s5d $^3D_2$ | 104.8559965 | 0.28755 | 0.30366 | 2.9063E+7 | 7.4 | 3.2411E+7 | 3.2 | 3.14E+07 | 7.9845E-3 | 8.9043E-3 |
| 3s3p $^3P_1$ | 3s5d $^3D_1$ | 104.8558316 | 0.16590 | 0.17519 | 1.6124E+7 | 7.9 | 1.7980E+7 | 2.7 | 1.75E+07 | 2.6578E-3 | 2.9637E-3 |
| 3s3p $^3P_1$ | 3s5d $^1D_2$ | 103.6672711 | 4.5776E-3 | 4.5998E-3 | 7.6216E+3 | | 7.6957E+3 | | | 2.0467E-6 | 2.0666E-6 |
| 3s3p $^3P_1$ | 3s7s $^3S_1$ | 98.9648474 | 0.071886 | 0.073043 | 3.6007E+6 | | 3.7176E+6 | | | 5.2872E-4 | 5.4587E-4 |
| 3s3p $^3P_1$ | 3s7s $^1S_0$ | 98.6712047 | 3.7813E-5 | 2.2757E-4 | 3.0156 | | 109.23 | | | 1.4673E-10 | 5.3144E-9 |
| 3s3p $^3P_1$ | 3s6d $^3D_1$ | 98.6573134 | 0.094612 | 0.10163 | 6.2958E+6 | -9.9 | 7.2644E+6 | 3.9 | 6.99E+06 | 9.1871E-4 | 1.0601E-4 |
| 3s3p $^3P_1$ | 3s6d $^3D_2$ | 98.6573134 | 0.16404 | 0.17621 | 1.1356E+7 | -9.2 | 1.3103E+7 | 4.8 | 1.25E+07 | 2.7618E-3 | 3.1868E-3 |
| 3s3p $^3P_1$ | 3s6d $^1D_2$ | 98.197834 | 3.1303E-3 | 3.1463E-3 | 4.1934E+3 | | 4.2364E+4 | | | 1.0104E-6 | 1.0207E-6 |
| 3s3p $^3P_1$ | 3s8s $^3S_1$ | 95.4845089 | 0.17817 | 0.17429 | 2.4627E+7 | | 2.3566E+7 | | | 3.3663E-3 | 3.2213E-3 |
| 3s3p $^3P_1$ | 3s8s $^1S_0$ | 95.3205702 | 2.3698E-4 | 1.4454E-4 | 131.38 | | 48.875 | | | 5.9656E-9 | 2.2193E-9 |
| 3s3p $^3P_1$ | 3s7d $^3D_1$ | 95.3183169 | 0.062369 | 0.068041 | 3.0336E+6 | | 3.6104E+6 | | | 4.1322E-4 | 4.9179E-4 |
| 3s3p $^3P_1$ | 3s7d $^3D_2$ | 95.3181806 | 0.10803 | 0.11775 | 5.4608E+6 | | 6.4877E+6 | | | 1.2397E-3 | 1.4729E-3 |
| 3s3p $^3P_1$ | 3s7d $^1D_2$ | 95.0973873 | 1.5787E-3 | 1.5842E-3 | 1.1743E+3 | | 1.1825E+3 | | | 2.6537E-7 | 2.6722E-7 |
| 3s3p $^3P_1$ | 3s9s $^3S_1$ | 93.3938425 | 0.17198 | 0.17097 | 2.4522E+7 | | 2.4234E+7 | | | 3.2067E-3 | 3.1691E-3 |
| 3s3p $^3P_1$ | 3s9s $^1S_0$ | 93.2920845 | 1.8382E-4 | 2.8813E-4 | 84.318 | | 207.16 | | | 3.6674E-9 | 9.0105E-9 |
| 3s3p $^3P_1$ | 3s8d $^3D_1$ | 93.2939471 | 0.049090 | 0.054199 | 2.0043E+6 | | 2.4433E+6 | | | 2.6155E-4 | 3.1882E-4 |
| 3s3p $^3P_1$ | 3s8d $^3D_2$ | 93.2937904 | 0.085439 | 0.094356 | 3.6429E+6 | | 4.4430E+6 | | | 7.9228E-4 | 9.6628E-4 |
| 3s3p $^3P_1$ | 3s8d $^1D_2$ | 93.1720537 | 6.7521E-3 | 7.7044E-3 | 2.2841E+5 | | 2.9738E+5 | | | 4.9546E-6 | 6.4507E-6 |
| 3s3p $^3P_2$ | $3p^2$ $^1D_2$ | 208.7527524 | 0.026063 | 0.026915 | 3.0259E+4 | | 3.2269E+5 | | | 1.9769E-5 | 2.1083E-5 |
| 3s3p $^3P_2$ | 3s4s $^3S_1$ | 186.2311489 | 1.9900 | 1.9773 | 4.1409E+8 | 1.5 | 4.0882E+8 | 0.20 | 4.08E+8 | 0.12919 | 0.12754 |
| 3s3p $^3P_2$ | $3p^2$ $^3P_1$ | 176.7731719 | 2.0614 | 2.1053 | 5.1954E+8 | 1.3 | 5.4191E+8 | 5.6 | 5.13E+8 | 0.14604 | 0.15233 |
| 3s3p $^3P_2$ | $3p^2$ $^3P_2$ | 176.3951844 | 3.5524 | 3.6282 | 9.3171E+8 | 1.2 | 9.7190E+8 | 5.6 | 9.21E+8 | 0.43464 | 0.45338 |
| 3s3p $^3P_2$ | 3s3d $^3D_3$ | 172.4981685 | 4.5970 | 4.6538 | 1.1917E+9 | 0.99 | 1.2213E+9 | 3.5 | 1.18E+8 | 0.74427 | 0.76278 |
| 3s3p $^3P_2$ | 3s3d $^3D_2$ | 172.4949251 | 1.9461 | 1.9702 | 2.9902E+8 | 0.68 | 3.0647E+8 | 3.2 | 2.97E+8 | 0.13339 | 0.13671 |
| 3s3p $^3P_2$ | 3s3d $^3D_1$ | 172.492158 | 0.50627 | 0.51257 | 3.3729E+7 | | 3.4574E+7 | | 3.32E+7 | 9.0274E-3 | 9.2535E-3 |
| 3s3p $^3P_2$ | 3s3d $^1D_2$ | 137.9081322 | 0.011240 | 0.011474 | 1.9519E+4 | | 2.0340E+4 | | | 5.5656E-6 | 5.7997E-6 |
| 3s3p $^3P_2$ | 3s5s $^3S_1$ | 121.1898957 | 0.60488 | 0.59838 | 1.3883E+8 | 12 | 1.3586E+8 | 9.6 | 1.24E+8 | 0.018342 | 0.017950 |
| 3s3p $^3P_2$ | 3s4d $^3D_3$ | 119.1814001 | 0.90998 | 0.94221 | 1.4158E+8 | -4.3 | 1.5179E+8 | 2.6 | 1.48E+8 | 0.042211 | 0.045254 |
| 3s3p $^3P_2$ | 3s4d $^3D_2$ | 119.1808035 | 0.38445 | 0.39799 | 3.5380E+7 | -4.4 | 3.7916E+7 | 2.5 | 3.70E+7 | 7.5343E-3 | 8.07432E-3 |
| 3s3p $^3P_2$ | 3s4d $^3D_1$ | 119.1803348 | 0.098866 | 0.10233 | 3.8997E+6 | -6.3 | 4.1777E+6 | 0.43 | 4.16E+6 | 4.9826E-4 | 5.3379E-4 |
| 3s3p $^3P_2$ | 3s4d $^1D_2$ | 114.6574139 | 2.6496E-3 | 2.6816E-3 | 1.8873E+3 | | 1.9332E+3 | | | 3.7199E-7 | 3.8103E-7 |
| 3s3p $^3P_2$ | 3s6s $^3S_1$ | 105.6661016 | 0.33810 | 0.33385 | 6.5438E+7 | 12 | 6.3803E+7 | 9.1 | 5.85E+7 | 6.5724E-3 | 6.4082E-3 |

| Upper | Lower | λ (nm) | | | | | | | | | |
|---|---|---|---|---|---|---|---|---|---|---|---|
| 3s3p $^3P_2$ | 3s5d $^3D_1$ | 104.9922116 | 0.042594 | 0.045001 | 1.0587E+6 | -8.7 | 1.1817E+6 | 1.9 | 1.16E+6 | 1.0498E-4 | 1.1718E-4 |
| 3s3p $^3P_2$ | 3s5d $^3D_2$ | 104.992377 | 0.16508 | 0.17442 | 9.5414E+6 | -8.3 | 1.0652E+7 | 2.4 | 1.04E+7 | 1.5769E-3 | 1.7604E-3 |
| 3s3p $^3P_2$ | 3s5d $^3D_3$ | 104.9922778 | 0.39094 | 0.41309 | 3.8222E+7 | -8.3 | 4.2676E+7 | 2.3 | 4.17E+7 | 8.8436E-3 | 9.8742E-3 |
| 3s3p $^3P_2$ | 3s5d $^1D_2$ | 103.800575 | 9.8619E-4 | 1.0139E-3 | 352.39 | | 372.47 | | | 5.6923E-8 | 6.0167E-8 |
| 3s3p $^3P_2$ | 3s7s $^3S_1$ | 99.0863249 | 0.093120 | 0.094315 | 6.0199E+6 | | 6.1754E+6 | | | 5.3167E-4 | 5.4540E-4 |
| 3s3p $^3P_2$ | 3s6d $^3D_1$ | 98.7780366 | 0.025351 | 0.027247 | 4.5036E+5 | -2.7 | 5.2024E+5 | 9.9 | 4.63E+5 | 3.9527E-5 | 4.5661E-5 |
| 3s3p $^3P_2$ | 3s6d $^3D_2$ | 98.7780366 | 0.093962 | 0.10100 | 3.7121E+6 | -11 | 4.2890E+6 | 3.1 | 4.16E+6 | 5.4302E-4 | 6.2741E-4 |
| 3s3p $^3P_2$ | 3s6d $^3D_3$ | 98.7777927 | 0.22252 | 0.23921 | 1.4871E+7 | -11 | 1.7185E+7 | 2.9 | 1.67E+7 | 3.0454E-3 | 3.5194E-3 |
| 3s3p $^3P_2$ | 3s6d $^1D_2$ | 98.3174347 | 5.0996E-4 | 5.2232E-4 | 110.89 | | 116.33 | | | 1.6070E-8 | 1.6858E-8 |
| 3s3p $^3P_2$ | 3s8s $^3S_1$ | 95.5975876 | 0.23062 | 0.22582 | 4.1115E+7 | | 3.9421E+7 | | | 3.3800E-3 | 3.2408E-3 |
| 3s3p $^3P_2$ | 3s7d $^3D_1$ | 95.4310021 | 1.5908E-2 | 1.7385E-2 | 1.9666E+5 | | 2.3487E+5 | | | 1.6110E-5 | 1.924E-5 |
| 3s3p $^3P_2$ | 3s7d $^3D_2$ | 95.4308655 | 6.2005E-2 | 6.7698E-2 | 1.7926E+6 | | 2.1369E+6 | | | 2.4476E-4 | 2.9176E-4 |
| 3s3p $^3P_2$ | 3s7d $^3D_3$ | 95.4306014 | 0.14668 | 0.15993 | 7.1655E+6 | | 8.5186E+6 | | | 1.3697E-3 | 1.6283E-3 |
| 3s3p $^3P_2$ | 3s7d $^1D_2$ | 95.2095505 | 1.0193E-4 | 4.1798E-5 | 4.8782 | | 0.8209 | | | 6.6297E-10 | 1.1148E-10 |
| 3s3p $^3P_2$ | 3s9s $^3S_1$ | 93.5020209 | 0.22261 | 0.22116 | 4.0942E+7 | | 4.0411E+7 | | | 3.2199E-4 | 3.1780E-4 |
| 3s3p $^3P_2$ | 3s8d $^3D_1$ | 93.401894 | 1.2550E-2 | 1.3845E-2 | 1.3055E+5 | | 1.5888E+5 | | | 1.0245E-5 | 1.2468E-5 |
| 3s3p $^3P_2$ | 3s8d $^3D_2$ | 93.401737 | 4.8614E-2 | 5.3716E-2 | 1.1753E+6 | | 1.4350E+6 | | | 1.5372E-4 | 1.8768E-4 |
| 3s3p $^3P_2$ | 3s8d $^3D_3$ | 93.4015014 | 0.11683 | 0.12927 | 4.8486E+6 | | 5.9361E+6 | | | 8.8782E-4 | 1.0870E-4 |
| 3s3p $^3P_2$ | 3s8d $^1D_2$ | 93.2797186 | 4.7739E-3 | 5.3640E-3 | 1.8964E+4 | | 2.3942E+4 | | | 1.4843E-6 | 1.8740E-6 |
| 3s3p $^1P_1$ | $3p^2$ $^1D_2$ | 390.1779718 | 0.19164 | 0.21239 | 2.5054E+5 | 48 | 3.0774E+5 | -36 | 4.80E+5 | 9.5307E-4 | 1.1706E-3 |
| 3s3p $^1P_1$ | 3s4s $^3S_1$ | 318.2434995 | 9.7510E-3 | 9.7904E-3 | 1.9924E+3 | | 2.0085E+3 | | | 3.0252E-6 | 3.0497E-6 |
| 3s3p $^1P_1$ | $3p^2$ $^3P_0$ | 292.1163066 | 0.037681 | 0.038175 | 1.1541E+5 | | 1.1846E+5 | | | 4.9216E-5 | 5.0515E-5 |
| 3s3p $^1P_1$ | $3p^2$ $^3P_1$ | 291.583954 | 4.8922E-3 | 5.1287E-3 | 652.02 | | 716.59 | | | 8.3112E-7 | 9.1341E-7 |
| 3s3p $^1P_1$ | $3p^2$ $^3P_2$ | 290.556957 | 0.010143 | 0.010195 | 1.6996E+3 | | 1.7170E+3 | | | 3.5852E-6 | 3.6221E-6 |
| 3s3p $^1P_1$ | 3s4s $^1S_0$ | 281.7014089 | 1.9317 | 1.9317 | 3.3820E+8 | -5.2 | 3.3593E+8 | -5.9 | 3.57E+8 | 0.13412 | 0.13322 |
| 3s3p $^1P_1$ | 3s3d $^3D_2$ | 280.1238931 | 1.4158E-2 | 1.4044E-2 | 3.6953E+3 | | 3.6360E+3 | | | 7.2455E-6 | 7.1293E-6 |

| Upper | Lower | λ (nm) | | | | | | | | | |
|---|---|---|---|---|---|---|---|---|---|---|---|
| 3s3p $^1P_1$ | 3s3d $^3D_1$ | 280.1165957 | 8.6734E-3 | 8.8318E-3 | 2.3116E+3 | | 2.3968E+3 | | | 2.7193E-6 | 2.8195E-6 |
| 3s3p $^1P_1$ | 3s3d $^1D_2$ | 199.0532628 | 5.1922 | 5.2517 | 1.3851E+9 | 0.37 | 1.4171E+9 | 2.7 | 1.38E+9 | 1.3714 | 1.4030 |
| 3s3p $^1P_1$ | $3p^2$ $^1S_0$ | 193.1049558 | 1.8996 | 1.9521 | 1.0153E+9 | 2.4 | 1.0722E+9 | 3.1 | 1.04E+9 | 0.18921 | 0.19981 |
| 3s3p $^1P_1$ | 3s5s $^3S_1$ | 166.0001759 | 1.8578E-3 | 1.8265E-3 | 5.0959 | | 4.9256 | | | 2.1053E-7 | 2.0349E-7 |
| 3s3p $^1P_1$ | 3s5s $^1S_0$ | 162.5627431 | 0.56092 | 0.56067 | 1.4839E+8 | 13 | 1.4826E+8 | 13 | 1.31E+8 | 0.019597 | 0.019580 |
| 3s3p $^1P_1$ | 3s4d $^3D_2$ | 162.2536381 | 1.6704E-3 | 1.8480E-3 | 2.6470 | | 3.2398 | | | 1.7413E-7 | 2.1312E-7 |
| 3s3p $^1P_1$ | 3s4d $^3D_1$ | 162.2527693 | 2.0818E-4 | 2.4729E-4 | 6.8525 | | 9.6690 | | | 2.7046E-9 | 3.8163E-9 |
| 3s3p $^1P_1$ | 3s4d $^1D_2$ | 153.9832937 | 2.3966 | 2.4404 | 6.3748E+8 | -4.9 | 6.6100E+8 | 1.34 | 6.70E+8 | 0.37769 | 0.39162 |
| 3s3p $^1P_1$ | 3s6s $^3S_1$ | 138.1912151 | 8.7981E-4 | 9.1661E-4 | 198.10 | | 215.02 | | | 5.6717E-8 | 6.1561E-8 |
| 3s3p $^1P_1$ | 3s6s $^1S_0$ | 137.1241581 | 0.28504 | 0.28445 | 6.3846E+7 | 9.5 | 6.3582E+7 | 9.1 | 5.83E+7 | 5.9995E-3 | 5.9747E-3 |
| 3s3p $^1P_1$ | 3s5d $^3D_2$ | 137.0411553 | 4.9401E-3 | 6.4099E-4 | 3.8425E+3 | | 64.691 | | | 1.8032E-6 | 3.0358E-8 |
| 3s3p $^1P_1$ | 3s5d $^3D_1$ | 137.0408736 | 1.8305E-4 | 1.2525E-4 | 8.7930 | | 4.1167 | | | 2.4757E-9 | 1.1591E-9 |
| 3s3p $^1P_1$ | 3s5d $^1D_2$ | 135.0177244 | 1.1539 | 1.1815 | 2.1921E+8 | -7.5 | 2.2982E+8 | -3.0 | 2.37E+8 | 0.099853 | 0.10469 |
| 3s3p $^1P_1$ | 3s7s $^3S_1$ | 127.1490251 | 2.7353E-4 | 4.1805E-5 | 24.582 | | 0.5742 | | | 5.9582E-9 | 1.3918E-10 |
| 3s3p $^1P_1$ | 3s7s $^1S_0$ | 126.6647227 | 6.7818E-2 | 7.1042E-2 | 4.5855E+6 | | 5.0319E+6 | | | 3.6766E-4 | 4.0345E-4 |
| 3s3p $^1P_1$ | 3s6d $^3D_1$ | 126.6418322 | 1.7347E-4 | 1.5099E-4 | 10.006 | | 7.5807 | | | 2.4060E-9 | 1.8228E-9 |
| 3s3p $^1P_1$ | 3s6d $^3D_2$ | 126.6418322 | 3.4194E-4 | 4.6179E-4 | 23.327 | | 42.546 | | | 9.3484E-9 | 1.7050E-8 |
| 3s3p $^1P_1$ | 3s6d $^1D_2$ | 125.8857161 | 0.67733 | 0.69612 | 9.3190E+7 | -8.6 | 9.8432E+7 | -3.5 | 1.02E+8 | 0.036901 | 0.038977 |
| 3s3p $^1P_1$ | 3s8s $^3S_1$ | 121.4610501 | 5.3653E-4 | 7.6079E-4 | 108.50 | | 218.15 | | | 2.3998E-8 | 4.8251E-8 |
| 3s3p $^1P_1$ | 3s8s $^1S_0$ | 121.1959029 | 0.19394 | 0.19145 | 4.2809E+7 | | 4.1717E+7 | | | 0.0031424 | 0.0030622 |
| 3s3p $^1P_1$ | 3s7d $^3D_2$ | 121.19204 | 1.9164E-3 | 2.0550E-3 | 836.08 | | 961.39 | | | 3.0684E-7 | 3.5283E-7 |
| 3s3p $^1P_1$ | 3s7d $^1D_2$ | 120.8353346 | 0.45454 | 0.46832 | 4.7453E+7 | | 5.0373E+7 | | | 0.017313 | 0.018378 |
| 3s3p $^1P_1$ | 3s9s $^1S_0$ | 117.9354874 | 0.17902 | 0.18112 | 3.9586E+7 | | 4.0520E+7 | | | 2.7515E-3 | 2.8165E-3 |
| 3s3p $^1P_1$ | 3s8d $^3D_2$ | 117.9382135 | 3.4536E-2 | 3.5427E-2 | 2.9463E+5 | | 3.1003E+5 | | | 1.0240E-4 | 1.0775E-4 |
| 3s3p $^1P_1$ | 3s8d $^1D_2$ | 117.7437333 | 0.37096 | 0.38318 | 3.4162E+7 | | 3.6449E+7 | | | 0.011834 | 0.012627 |

**Table 3**. The transition amplitudes of the E1 transitions arising from 5 lowest energy states.

| Transition | | $\lambda$ (nm) | RMEs (a. u.) | Ref | $A_{ki}$ ($s^{-1}$) | Diff (%) | NISTDATA | $f_{ik}$ |
|---|---|---|---|---|---|---|---|---|
| Initial | Final | | | | | | | |
| $3s^2\,^1S_0$ | $3s7s\,^3S_1$ | 72.2022234 | 6.8000E-3 | | 1.1045 | | | 2.5898E-10 |
| $3s^2\,^1S_0$ | $3s8s\,^3S_1$ | 70.3319258 | 4.2200E-3 | | 0.46024 | | | 1.0239E-10 |
| $3s^2\,^1S_0$ | $3s7d\,^3D_1$ | 70.2417172 | 1.4700E-3 | | 0.056062 | | | 1.2440E-11 |
| $3s^2\,^1S_0$ | $3s9s\,^3S_1$ | 69.1910562 | 3.2000E-3 | | 0.27795 | | | 5.9848E-11 |
| $3s3p\,^3P_0$ | $3s3p\,^3P_1$ | 164257.55584 | 1.41418 | 1.41769[8] | 4.0574E-6 | | | 4.9236E-9 |
| | | | | 1.41784[8] | | | | |
| $3s3p\,^3P_0$ | $3s3p\,^1P_1$ | 445.25599 | 4.7700E-3 | | 2.3175E-3 | | | 2.0664E-11 |
| $3s3p\,^3P_0$ | $3s8p\,^1P_1$ | 94.7255219 | 1.5600E-3 | | 0.025743 | | | 1.0389E-11 |
| $3s3p\,^3P_0$ | $3s8p\,^3P_1$ | 94.5414329 | 3.3400E-3 | | 0.118670 | | | 4.7716E-11 |
| $3s3p\,^3P_1$ | $3s3p\,^3P_2$ | 80723.2805941 | 1.58108 | 1.5811[22] | 2.5638E-5 | 0.15 | 2.56E-5 | 4.1743E-9 |
| $3s3p\,^3P_1$ | $3s3p\,^1P_1$ | 446.466242 | 4.1700E-3 | | 1.7568E-3 | | | 5.2500E-12 |
| $3s3p\,^3P_1$ | $3s4p\,^3P_2$ | 147.0220247 | 1.4000E-3 | | 3.3272E-3 | | | 1.7970E-12 |
| $3s3p\,^3P_1$ | $3s7p\,^3P_0$ | 97.4316149 | 1.1200E-3 | | 0.036583 | | | 1.7354E-12 |
| $3s3p\,^3P_1$ | $3s8p\,^3P_0$ | 94.5970963 | 3.5500E-3 | | 0.40157 | | | 1.7958E-11 |
| $3s3p\,^3P_1$ | $3s7f\,^3F_2$ | 94.5073107 | 1.7900E-3 | | 0.020478 | | | 4.5700E-12 |
| $3s3p\,^3P_2$ | $3s3p\,^1P_1$ | 448.9493015 | 5.3000E-3 | | 2.7911E-3 | | | 5.0604E-12 |
| $3s3p\,^3P_2$ | $3s4p\,^3P_1$ | 147.3845216 | 1.4100E-3 | | 5.5834E-3 | | | 1.0910E-12 |
| $3s3p\,^3P_2$ | $3s4p\,^1P_1$ | 144.2049024 | 1.1200E-3 | | 3.7611E-3 | | | 7.0353E-13 |
| $3s3p\,^3P_2$ | $3s7p\,^1P_1$ | 97.7123678 | 1.0800E-3 | | 0.011241 | | | 9.6544E-13 |
| $3s3p\,^3P_2$ | $3s7p\,^3P_2$ | 124.6204062 | 1.3800E-3 | | 5.3083 | | | 1.2359E-12 |
| $3s3p\,^3P_2$ | $3s8p\,^1P_1$ | 94.8915963 | 3.2900E-3 | | 0.11390 | | | 9.2255E-12 |
| $3s3p\,^3P_2$ | $3s7f\,^3F_3$ | 94.6117922 | 1.2300E-3 | | 6.8836E-3 | | | 1.2933E-12 |
| $3s3p\,^3P_2$ | $3s7f\,^1F_3$ | 93.277787 | 1.2900E-3 | | 7.9010E-3 | | | 1.4429E-12 |
| $3s3p\,^1P_1$ | $3s4p\,^3P_0$ | 219.4161336 | 1.1400E-3 | | 3.3185E-3 | | | 7.9839E-13 |

| | | | | | |
|---|---|---|---|---|---|
| 3s3p $^1P_1$ | 3s4p $^3P_1$ | 219.3488497 | 1.0100E-3 | 8.691E-3 | 6.2688E-13 |
| 3s3p $^1P_1$ | 3s4p $^3P_2$ | 219.2073418 | 1.1700E-3 | 7.0109E-4 | 8.4177E-13 |
| 3s3p $^1P_1$ | 3s7p $^3P_1$ | 124.626308 | 2.0200E-3 | 0.018954 | 4.4133E-12 |
| 3s3p $^1P_1$ | 3s7p $^3P_2$ | 124.6204062 | 1.0800E-3 | 3.2512E-3 | 1.2616E-12 |
| 3s3p $^1P_1$ | 3s8p $^3P_2$ | 120.0220552 | 5.2600E-3 | 0.086329 | 3.1073E-11 |
| 3s3p $^1P_1$ | 3s7f $^3F_2$ | 119.8842254 | 1.7400E-3 | 9.4794E-3 | 3.4042E-12 |

**Table 4**. The transition amplitudes of the M1 transitions arising from 5 lowest energy states.

| Transition | | λ (nm) | RMEs (a. u.) | Ref | $A_{ki}$ ($s^{-1}$) | Diff (%) | NISTDATA | $f_{ik}$ |
|---|---|---|---|---|---|---|---|---|
| Initial | Final | | | | | | | |
| $3s^2$ $^1S_0$ | $3p^2$ $^1D_2$ | 116.984582 | 9.1872 | | 8.6284E+3 | 3.1 | 8.37E+3 | 8.8518E-6 |
| $3s^2$ $^1S_0$ | $3p^2$ $^3P_2$ | 106.0797711 | 0.086549 | | 1.2490 | | | 1.0536E-9 |
| $3s^2$ $^1S_0$ | 3s3d $^3D_2$ | 104.6566889 | 5.8255E-3 | | 6.0540E-3 | | | 4.9707E-12 |
| $3s^2$ $^1S_0$ | 3s3d $^1D_2$ | 90.8349118 | 5.5608 | | 1.1200E+4 | -3.4 | 1.16E+4 | 6.9273E-6 |
| $3s^2$ $^1S_0$ | 3s4d $^3D_2$ | 82.3154208 | 1.6056E-3 | | 1.5278E-3 | | | 7.7604E-13 |
| $3s^2$ $^1S_0$ | 3s4d $^1D_2$ | 80.1319741 | 3.5630 | | 8.6062E+3 | | | 4.1425E-6 |
| $3s^2$ $^1S_0$ | 3s5d $^3D_2$ | 75.2882788 | 1.4406E-3 | | 1.9216E-3 | | | 8.1650E-13 |
| $3s^2$ $^1S_0$ | 3s5d $^1D_2$ | 75.2881937 | 1.8119 | | 3.0398E+3 | | | 1.2916E-6 |
| $3s^2$ $^1S_0$ | 3s6d $^3D_2$ | 72.0383918 | 1.2367E-3 | | 1.7657E-3 | | | 6.8689E-13 |
| $3s^2$ $^1S_0$ | 3s6d $^1D_2$ | 71.7931008 | 1.0502 | | 1.2952E+3 | | | 5.0043E-7 |
| $3s^2$ $^1S_0$ | 3s7d $^3D_2$ | 70.2416431 | 2.3001E-3 | | 6.9299E-3 | | | 2.5631E-12 |
| $3s^2$ $^1S_0$ | 3s7d $^1D_2$ | 70.1216688 | 0.67362 | | 599.48 | | | 2.2097E-7 |
| $3s^2$ $^1S_0$ | 3s8d $^3D_2$ | 69.1361261 | 4.9249E-2 | | 3.4394 | | | 1.2323E-9 |
| $3s^2$ $^1S_0$ | 3s8d $^1D_2$ | 69.0692497 | 0.53215 | | 403.51 | | | 1.4430E-7 |
| 3s3p $^3P_0$ | 3s3p $^3P_2$ | 54124.2693223 | 6.2721 | | 1.8970E-10 | | | 4.1658E-14 |
| 3s3p $^3P_0$ | 3s4p $^3P_2$ | 146.8905474 | 3.9900 | | 521.42 | | | 8.4336E-7 |
| 3s3p $^3P_0$ | 3s4f $^3F_2$ | 116.2446694 | 6.6591 | | 4.6792E+3 | | | 4.7398E-6 |
| 3s3p $^3P_0$ | 3s5p $^3P_2$ | 113.2135239 | 1.5191 | | 277.90 | | | 2.6701E-7 |
| 3s3p $^3P_0$ | 3s5f $^3F_2$ | 104.1182083 | 4.8853 | | 4.36876E+3 | | | 3.5502E-6 |
| 3s3p $^3P_0$ | 3s6p $^3P_2$ | 102.4284767 | 0.92101 | | 168.51 | | | 1.3253E-7 |
| 3s3p $^3P_0$ | 3s6f $^3F_2$ | 98.8842201 | 4.5173 | | 4.8343E+3 | | | 3.5435E-6 |
| 3s3p $^3P_0$ | 3s7p $^3P_2$ | 97.368452 | 0.65536 | | 109.92 | | | 7.8119E-8 |
| 3s3p $^3P_0$ | 3s6d $^3F_2$ | 96.4395855 | 2.8996 | | 2.2574E+3 | | | 1.5738E-6 |
| 3s3p $^3P_0$ | 3s7p $^3P_2$ | 94.5385013 | 0.15642 | | 7.2569 | | | 4.8619E-9 |

| | | | | | |
|---|---|---|---|---|---|
| 3s3p $^3P_0$ | 3s6f $^3F_2$ | 94.4529662 | 0.52466 | 82.014 | 5.4848E-8 |
| 3s3p $^3P_1$ | 3s3p $^3P_2$ | 80723.2805941 | 9.4123 | 5.7890E-11 | 9.4259E-15 |
| 3s3p $^3P_1$ | 3s3p $^1P_1$ | 446.466242 | 0.045467 | 4.3501E-4 | 1.3000E-12 |
| 3s3p $^3P_1$ | 3s4p $^3P_1$ | 147.0856666 | 3.46890 | 652.51 | 2.1164E-7 |
| 3s3p $^3P_1$ | 3s4p $^3P_2$ | 147.0220247 | 5.9957 | 1.1721E+3 | 6.3309E-7 |
| 3s3p $^3P_1$ | 3s4p $^1P_1$ | 143.9539692 | 0.060143 | 0.21843 | 6.7862E-11 |
| 3s3p $^3P_1$ | 3s4f $^3F_2$ | 116.3269937 | 6.6655 | 4.6717E+3 | 1.5796E-6 |
| 3s3p $^3P_1$ | 3s4f $^3F_3$ | 116.324328 | 9.41720 | 6.6615E+3 | 3.1533E-6 |
| 3s3p $^3P_1$ | 3s4f $^1F_3$ | 116.256643 | 0.32217 | 7.8192 | 3.6970E-09 |
| 3s3p $^3P_1$ | 3s5p $^3P_1$ | 113.3081305 | 1.3160 | 346.15 | 6.6628E-8 |
| 3s3p $^3P_1$ | 3s5p $^3P_2$ | 113.2916095 | 2.2810 | 624.41 | 2.0026E-7 |
| 3s3p $^3P_1$ | 3s5p $^1P_1$ | 113.1028395 | 0.078938 | 1.2568 | 2.4104E-10 |
| 3s3p $^3P_1$ | 3s5f $^3F_2$ | 104.1842477 | 4.8838 | 4.3523E+3 | 1.1804E-6 |
| 3s3p $^3P_1$ | 3s5f $^3F_3$ | 104.1784192 | 6.8986 | 6.2046E+3 | 2.3557E-6 |
| 3s3p $^3P_1$ | 3s5f $^1F_3$ | 103.9199973 | 0.085948 | 0.97512 | 3.6839E-10 |
| 3s3p $^3P_1$ | 3s6p $^1P_1$ | 102.6004178 | 0.030154 | 0.29854 | 4.7116E-11 |
| 3s3p $^3P_1$ | 3s6p $^3P_1$ | 102.6004178 | 0.79788 | 209.02 | 3.2988E-8 |
| 3s3p $^3P_1$ | 3s6p $^3P_2$ | 102.4923894 | 1.3820 | 378.24 | 9.9282E-8 |
| 3s3p $^3P_1$ | 3s6f $^3F_2$ | 98.9437849 | 4.5055 | 4.7946E+3 | 1.1729E-6 |
| 3s3p $^3P_1$ | 3s6f $^3F_3$ | 98.9266556 | 6.37440 | 6.8611E+3 | 2.3489E-6 |
| 3s3p $^3P_1$ | 3s6f $^1F_3$ | 98.2400974 | 0.040599 | 0.28818 | 9.7296E-11 |
| 3s3p $^3P_1$ | 3s7p $^1P_1$ | 97.5942338 | 0.017600 | 0.13061 | 1.8650E-11 |
| 3s3p $^3P_1$ | 3s7p $^3P_1$ | 97.4298113 | 0.56464 | 135.56 | 1.9293E-8 |
| 3s3p $^3P_1$ | 3s7p $^3P_2$ | 97.4262042 | 0.98276 | 246.45 | 5.8452E-8 |
| 3s3p $^3P_1$ | 3p3d $^3F_2$ | 96.4962407 | 2.8808 | 2.2217E+3 | 5.1692E-7 |
| 3s3p $^3P_1$ | 3p3d $^3F_3$ | 96.4728093 | 4.16820 | 3.3262E+3 | 1.0830E-6 |
| 3s3p $^3P_1$ | 3s8p $^1P_1$ | 94.7801805 | 0.18260 | 16.273 | 2.1917E-9 |
| 3s3p $^3P_1$ | 3s7f $^1F_3$ | 94.5010321 | 0.019136 | 0.077732 | 2.4284E-11 |
| 3s3p $^3P_1$ | 3s8p $^3P_1$ | 94.5958793 | 0.40449 | 80.633 | 1.0818E-8 |
| 3s3p $^3P_1$ | 3s8p $^3P_2$ | 94.5929443 | 0.78789 | 183.59 | 4.1047E-8 |
| 3s3p $^3P_1$ | 3s7f $^3F_2$ | 94.5073107 | 0.025766 | 0.19723 | 4.4018E-11 |
| 3s3p $^3P_1$ | 3s7f $^3F_3$ | 94.5010321 | 0.85032 | 153.48 | 4.7950E-8 |
| 3s3p $^3P_1$ | 3s9p $^1P_1$ | 93.034342 | 0.025194 | 0.33997 | 4.4116E-11 |
| 3s3p $^3P_2$ | 3s3p $^1P_1$ | 448.9493015 | 0.038705 | 3.0662E-4 | 5.5593E-13 |
| 3s3p $^3P_2$ | 3s4p $^3P_0$ | 147.3845216 | 4.0252 | 2.6091E+3 | 1.6994E-7 |
| 3s3p $^3P_2$ | 3s4p $^3P_1$ | 147.3541602 | 6.0307 | 1.9543E+3 | 3.8171E-7 |
| 3s3p $^3P_2$ | 3s4p $^3P_2$ | 147.2902858 | 5.3072 | 910.06 | 2.9600E-7 |
| 3s3p $^3P_2$ | 3s4p $^1P_1$ | 144.2111412 | 0.077539 | 0.35984 | 6.7317E-11 |
| 3s3p $^3P_2$ | 3s4f $^3F_2$ | 116.4948697 | 2.5255 | 665.84 | 1.3547E-7 |
| 3s3p $^3P_2$ | 3s4f $^3F_3$ | 116.4921963 | 6.6719 | 3.3197E+3 | 9.4556E-7 |
| 3s3p $^3P_2$ | 3s4f $^3F_4$ | 116.4882474 | 13.104 | 9.9617E+3 | 3.6479E-6 |
| 3s3p $^3P_2$ | 3s4f $^1F_3$ | 116.4243158 | 0.21488 | 3.4535 | 9.8252E-10 |
| 3s3p $^3P_2$ | 3s5p $^3P_0$ | 113.4747266 | 1.5260 | 1.3861E+3 | 5.3517E-8 |
| 3s3p $^3P_2$ | 3s5p $^3P_1$ | 113.4674003 | 0.22847 | 10.360 | 1.1999E-9 |
| 3s3p $^3P_2$ | 3s5p $^3P_2$ | 113.4508328 | 2.0165 | 484.58 | 9.3509E-8 |
| 3s3p $^3P_2$ | 3s5p $^1P_1$ | 113.2615323 | 0.12537 | 3.1480 | 3.6326E-10 |
| 3s3p $^3P_2$ | 3s5f $^3F_2$ | 104.3188853 | 1.8447 | 616.94 | 1.0066E-7 |
| 3s3p $^3P_2$ | 3s5f $^3F_3$ | 104.3130417 | 4.8751 | 3.0786E+3 | 7.0312E-7 |
| 3s3p $^3P_2$ | 3s5f $^3F_4$ | 104.3054363 | 9.5616 | 9.2143E+3 | 2.7053E-6 |
| 3s3p $^3P_2$ | 3s5f $^1F_3$ | 104.0539523 | 0.056890 | 0.42448 | 9.6467E-11 |
| 3s3p $^3P_2$ | 3s6p $^1P_1$ | 102.7309903 | 0.060857 | 1.2083 | 1.1471E-10 |
| 3s3p $^3P_2$ | 3s6p $^3P_0$ | 102.6330509 | 0.92275 | 837.36 | 2.6448E-8 |
| 3s3p $^3P_2$ | 3s6p $^3P_1$ | 102.6294591 | 1.3824 | 626.56 | 5.9365E-8 |
| 3s3p $^3P_2$ | 3s6p $^3P_2$ | 102.6226869 | 1.2198 | 292.80 | 4.6230E-8 |
| 3s3p $^3P_2$ | 3s6f $^3F_2$ | 99.0652107 | 1.6933 | 673.09 | 9.9034E-8 |
| 3s3p $^3P_2$ | 3s6f $^3F_3$ | 99.0480393 | 4.4829 | 3372.64 | 6.9448E-7 |
| 3s3p $^3P_2$ | 3s6f $^3F_4$ | 99.0255784 | 8.8089 | 1.0140E+4 | 2.6834E-06 |
| 3s3p $^3P_2$ | 3s6f $^1F_3$ | 98.3598011 | 0.029847 | 0.15481 | 3.1436E-11 |
| 3s3p $^3P_2$ | 3s7p $^1P_1$ | 97.7123678 | 0.037994 | 0.60498 | 5.1959E-11 |
| 3s3p $^3P_2$ | 3s7p $^3P_0$ | 97.5493553 | 0.65011 | 535.83 | 1.5289E-8 |
| 3s3p $^3P_2$ | 3s7p $^3P_1$ | 97.5475473 | 0.97673 | 403.20 | 3.4513E-8 |
| 3s3p $^3P_2$ | 3s7p $^3P_2$ | 97.5439315 | 0.86622 | 190.31 | 2.7148E-8 |
| 3s3p $^3P_2$ | 3p3d $^3F_2$ | 96.61173 | 1.0721 | 305.87 | 4.2802E-8 |
| 3s3p $^3P_2$ | 3p3d $^3F_3$ | 96.5882424 | 2.9050 | 1.6060E+3 | 3.1449E-7 |
| 3s3p $^3P_2$ | 3p3d $^3F_4$ | 96.5574749 | 5.8762 | 5.1192E+3 | 1.2880E-6 |
| 3s3p $^3P_2$ | 3p7f $^1F_3$ | 95.2142825 | 0.018339 | 0.068758 | 1.3084E-11 |
| 3s3p $^3P_2$ | 3s8p $^3P_0$ | 94.7080817 | 0.063337 | 5.8960 | 1.5858E-10 |
| 3s3p $^3P_2$ | 3s8p $^3P_1$ | 94.7068619 | 0.32171 | 50.708 | 4.0913E-9 |
| 3s3p $^3P_2$ | 3s8p $^3P_2$ | 94.70392 | 0.50509 | 75.008 | 1.0086E-8 |
| 3s3p $^3P_2$ | 3s7f $^3F_2$ | 94.6180854 | 0.19800 | 11.579 | 1.5541E-9 |
| 3s3p $^3P_2$ | 3s7f $^3F_3$ | 94.6117922 | 0.56877 | 68.270 | 1.2827E-8 |
| 3s3p $^3P_2$ | 3s7f $^3F_4$ | 94.6020451 | 1.2514 | 257.17 | 6.2111E-8 |
| 3s3p $^3P_2$ | 3s8f $^1F_3$ | 94.6955259 | 0.028597 | 3.0662E-4 | 5.5593E-13 |
| 3s3p $^1P_1$ | 3s4p $^3P_1$ | 219.3488497 | 0.18773 | 0.25909 | 1.8689E-10 |
| 3s3p $^1P_1$ | 3s4p $^3P_2$ | 219.2073418 | 0.041882 | 7.7623E-3 | 9.3201E-12 |
| 3s3p $^1P_1$ | 3s4p $^1P_1$ | 212.4561331 | 10.450 | 941.77 | 6.3732E-7 |
| 3s3p $^1P_1$ | 3s4f $^3F_2$ | 157.3156661 | 0.033023 | 0.025350 | 1.5676E-11 |
| 3s3p $^1P_1$ | 3s4f $^3F_3$ | 157.3107909 | 0.50700 | 4.2688 | 3.6955E-9 |
| 3s3p $^1P_1$ | 3s4f $^1F_3$ | 157.1871547 | 13.949 | 3.2440E+3 | 2.8039E-6 |
| 3s3p $^1P_1$ | 3s5p $^3P_1$ | 151.8445852 | 0.16851 | 1.3131 | 4.5392E-10 |
| 3s3p $^1P_1$ | 3s5p $^3P_2$ | 151.8149169 | 0.010228 | 2.9055E-3 | 1.6733E-12 |
| 3s3p $^1P_1$ | 3s5p $^1P_1$ | 151.4761349 | 2.9501 | 407.39 | 1.4014E-7 |
| 3s3p $^1P_1$ | 3s5f $^3F_2$ | 135.8959873 | 2.6584E-3 | 3.4152E-4 | 1.5760E-13 |
| 3s3p $^1P_1$ | 3s5f $^3F_3$ | 135.8860709 | 0.10444 | 0.37665 | 2.4330E-10 |
| 3s3p $^1P_1$ | 3s5f $^1F_3$ | 135.4467358 | 6.5165 | 1.4903E+3 | 9.5643E-7 |
| 3s3p $^1P_1$ | 3s6p $^1P_1$ | 133.2136541 | 1.5537 | 214.81 | 5.7150E-8 |
| 3s3p $^1P_1$ | 3s6p $^3P_1$ | 133.0429803 | 0.066375 | 0.39455 | 1.0470E-10 |
| 3s3p $^1P_1$ | 3s6p $^3P_2$ | 133.0315999 | 3.0111E-3 | 4.8740E-4 | 2.1553E-13 |
| 3s3p $^1P_1$ | 3s6f $^3F_2$ | 127.1142597 | 0.065350 | 0.28822 | 1.1637E-10 |
| 3s3p $^1P_1$ | 3s6f $^3F_3$ | 127.0859894 | 0.025823 | 0.032181 | 1.8182E-11 |
| 3s3p $^1P_1$ | 3s6f $^1F_3$ | 125.9551811 | 3.6888 | 686.71 | 3.8111E-07 |
| 3s3p $^1P_1$ | 3s7p $^1P_1$ | 124.8954624 | 1.0127 | 125.98 | 2.9461E-08 |
| 3s3p $^1P_1$ | 3s7p $^3P_1$ | 124.626308 | 0.043526 | 0.23524 | 5.4777E-11 |
| 3s3p $^1P_1$ | 3s7p $^3P_2$ | 124.6204062 | 0.040003 | 0.11925 | 4.6275E-11 |
| 3s3p $^1P_1$ | 3s3d $^3F_2$ | 123.1028769 | 0.24638 | 4.8093 | 1.8211E-9 |
| 3s3p $^1P_1$ | 3s3d $^3F_3$ | 123.0647453 | 0.021818 | 0.026980 | 1.4294E-11 |
| 3s3p $^1P_1$ | 3s7f $^1F_3$ | 120.8429569 | 2.3209 | 334.42 | 1.7084E-7 |
| 3s3p $^1P_1$ | 3s8p $^1P_1$ | 120.3236513 | 0.32749 | 15.874 | 3.4457E-9 |
| 3s3p $^1P_1$ | 3s7f $^3F_2$ | 119.8842254 | 0.070468 | 0.44914 | 1.6130E-10 |
| 3s3p $^1P_1$ | 3s7f $^3F_3$ | 119.8741225 | 0.046541 | 0.14000 | 7.0376E-11 |
| 3s3p $^1P_1$ | 3s8f $^1F_3$ | 117.7406557 | 1.6878 | 201.41 | 9.7677E-8 |

**Table 5**. The transition amplitudes of the E2 transitions arising from 5 lowest energy states.

| Transition | | λ (nm) | RMEs (a. u.) | Ref | $A_{ki}$ ($s^{-1}$) | Diff (%) | NISTDATA | $f_{ik}$ |
|---|---|---|---|---|---|---|---|---|
| Initial | Final | | | | | | | |
| $3s^2$ $^1S_0$ | $3s3p$ $^3P_2$ | 266.1146384 | 12.427 | | 3.4505E-3 | 3.0 | 3.35E-3 | 1.8317E-11 |
| $3s^2$ $^1S_0$ | $3s4p$ $^3P_2$ | 94.812855 | 0.59500 | | 1.3778E-3 | | | 9.2845E-13 |
| $3s^2$ $^1S_0$ | $3s5p$ $^3P_2$ | 79.5407634 | 0.42839 | | 1.7188E-3 | | | 8.1515E-13 |
| $3s^2$ $^1S_0$ | $3s5f$ $^3F_2$ | 74.941334 | 1.9243E-3 | | 4.6713E-8 | | | 1.9666E-17 |
| $3s^2$ $^1S_0$ | $3s6p$ $^3P_2$ | 74.0619334 | 0.31573 | | 1.3340E-3 | | | 5.4850E-13 |
| $3s^2$ $^1S_0$ | $3s6f$ $^3F_2$ | 72.1910116 | 2.0305E-3 | | 6.2703E-8 | | | 2.4495E-17 |
| $3s^2$ $^1S_0$ | $3s7p$ $^3P_2$ | 71.3797782 | 0.30790 | | 1.5256E-3 | | | 5.8267E-13 |
| $3s^2$ $^1S_0$ | $3s3d$ $^3F_2$ | 70.8793124 | 5.9542E-3 | | 5.9095E-7 | | | 2.2254E-16 |
| $3s^2$ $^1S_0$ | $3s8p$ $^3P_2$ | 69.8470154 | 0.12093 | | 2.6232E-4 | | | 9.5930E-14 |
| $3s3p$ $^3P_0$ | $3p^2$ $^1D_2$ | 207.950704 | 9.0209 | | 6.2402E-3 | | | 2.0228E-11 |
| $3s3p$ $^3P_0$ | $3p^2$ $^3P_2$ | 175.8221664 | 4.5555 | | 3.6830E-3 | | | 8.5345E-12 |
| $3s3p$ $^3P_0$ | $3s3d$ $^3D_2$ | 171.9469275 | 2.0328 | | 8.1982E-4 | | | 1.8169E-12 |
| $3s3p$ $^3P_0$ | $3s3d$ $^1D_2$ | 137.5576366 | 0.21739 | | 2.8612E-5 | | | 4.0584E-14 |
| $3s3p$ $^3P_0$ | $3s4d$ $^3D_2$ | 118.9189459 | 0.37575 | | 1.7703E-4 | | | 1.8766E-13 |
| $3s3p$ $^3P_0$ | $3s4d$ $^1D_2$ | 118.9184792 | 0.41312 | | 2.1400E-4 | | | 2.2685E-13 |
| $3s3p$ $^3P_0$ | $3s5d$ $^3D_2$ | 104.789103 | 0.14915 | | 5.2501E-5 | | | 4.3214E-14 |
| $3s3p$ $^3P_0$ | $3s5d$ $^1D_2$ | 104.7889383 | 0.26625 | | 1.6730E-4 | | | 1.3771E-13 |
| $3s3p$ $^3P_0$ | $3s6d$ $^3D_2$ | 98.5980928 | 0.087828 | | 2.4684E-5 | | | 1.7988E-14 |
| $3s3p$ $^3P_0$ | $3s6d$ $^1D_2$ | 98.1391636 | 0.16211 | | 8.6081E-5 | | | 6.2147E-14 |
| $3s3p$ $^3P_0$ | $3s7d$ $^3D_2$ | 95.2628998 | 0.017049 | | 1.1048E-6 | | | 7.5154E-16 |
| $3s3p$ $^3P_0$ | $3s7d$ $^1D_2$ | 95.0423622 | 0.10571 | | 4.2968E-5 | | | 2.9094E-14 |
| $3s3p$ $^3P_0$ | $3s8d$ $^3D_2$ | 93.2408322 | 0.10993 | | 5.1133E-5 | | | 3.3323E-14 |
| $3s3p$ $^3P_0$ | $3s8d$ $^1D_2$ | 93.1192335 | 0.093575 | | 3.7293E-5 | | | 2.4240E-14 |
| $3s3p$ $^3P_1$ | $3s4s$ $^3S_1$ | 185.8024967 | 3.4413 | | 2.6579E-3 | | | 1.3756E-12 |

| | | | | | |
|---|---|---|---|---|---|
| 3s3p $^3P_1$ | 3p$^2$ $^3P_1$ | 176.3869082 | 7.6426 | 0.017002 | 7.9303E-12 |
| 3s3p $^3P_1$ | 3s3d $^3D_3$ | 172.1303419 | 2.4665 | 8.5752E-4 | 8.8879E-13 |
| 3s3p $^3P_1$ | 3s3d $^3D_1$ | 172.124357 | 2.0144 | 1.3348E-3 | 5.9289E-13 |
| 3s3p $^3P_1$ | 3s5s $^3S_1$ | 121.008226 | 1.0222 | 2.0015E-3 | 4.3937E-13 |
| 3s3p $^3P_1$ | 3s4d $^3D_3$ | 118.9184792 | 0.45464 | 1.8512E-3 | 9.1579E-14 |
| 3s3p $^3P_1$ | 3s4d $^3D_1$ | 119.0056978 | 0.45140 | 4.2426E-4 | 9.0080E-14 |
| 3s3p $^3P_1$ | 3s6s $^3S_1$ | 105.5279664 | 0.56897 | 1.2294E-3 | 2.0525E-13 |
| 3s3p $^3P_1$ | 3s5d $^3D_1$ | 104.8558316 | 0.21652 | 1.8381E-4 | 3.0299E-14 |
| 3s3p $^3P_1$ | 3s5d $^3D_3$ | 104.8558975 | 0.18155 | 5.5386E-5 | 2.1302E-14 |
| 3s3p $^3P_1$ | 3s7s $^3S_1$ | 98.9648474 | 0.22250 | 2.5918E-4 | 3.8056E-14 |
| 3s3p $^3P_1$ | 3s6d $^3D_1$ | 98.6573134 | 0.12819 | 8.7380E-5 | 1.2751E-14 |
| 3s3p $^3P_1$ | 3s6d $^3D_3$ | 98.65707 | 0.099923 | 2.2754E-5 | 7.7474E-15 |
| 3s3p $^3P_1$ | 3s6g $^3G_3$ | 97.90753 | 4.1249E-3 | 4.0283E-8 | 1.3508E-17 |
| 3s3p $^3P_1$ | 3s8s $^3S_1$ | 95.4845089 | 0.34692 | 7.5360E-4 | 1.0301E-13 |
| 3s3p $^3P_1$ | 3s7d $^3D_1$ | 95.3183169 | 0.14572 | 1.3412E-4 | 1.8269E-14 |
| 3s3p $^3P_1$ | 3s7d $^3D_3$ | 95.3179171 | 0.025384 | 1.7443E-6 | 5.5438E-16 |
| 3s3p $^3P_1$ | 3s7g $^3G_3$ | 94.8787397 | 0.031900 | 2.8191E-6 | 8.8773E-16 |
| 3s3p $^3P_1$ | 3s9s $^3S_1$ | 93.3938425 | 0.40483 | 1.1463E-3 | 1.4990E-13 |
| 3s3p $^3P_1$ | 3s8d $^3D_1$ | 93.2939471 | 0.016335 | 1.8764E-6 | 2.4484E-16 |
| 3s3p $^3P_1$ | 3s8d $^3D_3$ | 93.2935554 | 0.086315 | 2.2454E-5 | 6.8364E-15 |
| 3s3p $^3P_2$ | 3p$^2$ $^1D_2$ | 208.7527524 | 11.822 | 0.010513 | 6.8682E-12 |
| 3s3p $^3P_2$ | 3p$^2$ $^3P_0$ | 176.9686926 | 4.6325 | 0.018434 | 1.7310E-12 |
| 3s3p $^3P_2$ | 3p$^2$ $^3P_2$ | 176.3951844 | 10.644 | 0.019782 | 9.2280E-12 |
| 3s3p $^3P_2$ | 3s4s $^1S_0$ | 173.0918056 | 4.2854 | 0.017623 | 1.5831E-12 |
| 3s3p $^3P_2$ | 3s3d $^3D_2$ | 172.4949251 | 6.2966 | 7.7416E-3 | 3.4534E-12 |
| 3s3p $^3P_2$ | 3s3d $^1D_2$ | 137.9081322 | 0.32423 | 6.2843E-5 | 1.7918E-14 |
| 3s3p $^3P_2$ | 3p$^2$ $^1S_0$ | 135.0264935 | 5.0716 | 0.085441 | 4.6708E-12 |
| 3s3p $^3P_2$ | 3s5s $^1S_0$ | 119.3474938 | 1.3215 | 0.010753 | 4.5925E-13 |
| 3s3p $^3P_2$ | 3s4d $^3D_2$ | 119.1808035 | 1.3170 | 2.1510E-3 | 4.5805E-13 |
| 3s3p $^3P_2$ | 3s4d $^1D_2$ | 114.6574139 | 0.53535 | 4.3128E-4 | 8.5002E-14 |
| 3s3p $^3P_2$ | 3s6s $^1S_0$ | 105.0410899 | 0.80448 | 7.5458E-3 | 2.4964E-13 |
| 3s3p $^3P_2$ | 3s5d $^3D_2$ | 104.992377 | 0.59715 | 8.3345E-4 | 1.3774E-13 |
| 3s3p $^3P_2$ | 3s5d $^1D_2$ | 103.800575 | 0.34531 | 2.9507E-4 | 4.7663E-14 |
| 3s3p $^3P_2$ | 3s7s $^1S_0$ | 98.791962 | 0.30051 | 1.4308E-3 | 4.1871E-14 |
| 3s3p $^3P_2$ | 3s6d $^3D_2$ | 98.7780366 | 0.36042 | 4.1192E-4 | 6.0255E-14 |
| 3s3p $^3P_2$ | 3s6d $^1D_2$ | 98.3174347 | 0.21547 | 1.5070E-4 | 2.1839E-14 |
| 3s3p $^3P_2$ | 3s8s $^1S_0$ | 95.4332608 | 0.52424 | 5.1764E-3 | 1.4136E-13 |
| 3s3p $^3P_2$ | 3s6g $^3G_4$ | 98.0264241 | 4.6864E-3 | 4.0197E-8 | 1.0423E-17 |
| 3s3p $^3P_2$ | 3s6g $^1G_4$ | 98.0264241 | 7.4147E-3 | 1.0062E-7 | 2.6093E-17 |
| 3s3p $^3P_2$ | 3s7d $^3D_2$ | 95.4308655 | 0.22025 | 1.8276E-4 | 2.4953E-14 |
| 3s3p $^3P_2$ | 3s7d $^1D_2$ | 95.2095505 | 0.10573 | 4.2608E-5 | 5.7905E-15 |
| 3s3p $^3P_2$ | 3s7g $^3G_4$ | 94.9903874 | 0.068274 | 9.9848E-6 | 2.4313E-15 |
| 3s3p $^3P_2$ | 3s7g $^1G_4$ | 94.9903874 | 1.3218E-3 | 3.7425E-9 | 9.1128E-19 |
| 3s3p $^3P_2$ | 3s9s $^1S_0$ | 93.401894 | 0.60191 | 7.5990E-3 | 1.9877E-13 |
| 3s3p $^3P_2$ | 3s8d $^3D_2$ | 93.401737 | 0.21864 | 2.0053E-4 | 2.6227E-14 |
| 3s3p $^3P_2$ | 3s8d $^1D_2$ | 93.2797186 | 0.16458 | 1.1437E-4 | 1.4919E-14 |
| 3s3p $^3P_2$ | 3s8g $^3G_4$ | 93.1191382 | 0.096989 | 2.2258E-5 | 5.2082E-15 |
| 3s3p $^1P_1$ | 3s4s $^3S_1$ | 318.2434995 | 6.9749 | 7.4067E-4 | 1.1246E-12 |
| 3s3p $^1P_1$ | 3p$^2$ $^3P_1$ | 291.583954 | 6.3860 | 9.6158E-4 | 1.2257E-12 |
| 3s3p $^1P_1$ | 3s3d $^3D_3$ | 280.1324466 | 14.906 | 2.7433E-3 | 7.5308E-12 |
| 3s3p $^1P_1$ | 3s3d $^3D_1$ | 280.1165957 | 1.6312 | 7.6678E-5 | 9.0200E-14 |
| 3s3p $^1P_1$ | 3s5s $^3S_1$ | 166.0001759 | 1.3613 | 7.3066E-4 | 3.0185E-13 |
| 3s3p $^1P_1$ | 3s4d $^3D_3$ | 162.2547438 | 1.6100 | 4.9095E-4 | 4.5213E-13 |
| 3s3p $^1P_1$ | 3s4d $^3D_1$ | 162.2527693 | 0.17504 | 1.3541E-5 | 5.3445E-15 |
| 3s3p $^1P_1$ | 3s6s $^3S_1$ | 138.1912151 | 0.69420 | 4.7524E-4 | 1.3606E-13 |
| 3s3p $^1P_1$ | 3s5d $^3D_3$ | 137.0409863 | 0.54565 | 1.3120E-4 | 8.6195E-14 |
| 3s3p $^1P_1$ | 3s5d $^3D_1$ | 137.0408736 | 0.059527 | 3.6436E-6 | 1.0259E-15 |
| 3s3p $^1P_1$ | 3s7s $^3S_1$ | 127.1490251 | 0.33052 | 1.6337E-4 | 3.9597E-14 |
| 3s3p $^1P_1$ | 3s6d $^3D_3$ | 126.6414312 | 0.28217 | 5.2061E-5 | 2.9208E-14 |
| 3s3p $^1P_1$ | 3s6d $^3D_1$ | 126.6418322 | 0.021897 | 7.3153E-7 | 1.7589E-16 |
| 3s3p $^1P_1$ | 3s8s $^3S_1$ | 121.4610501 | 0.36677 | 2.5290E-4 | 5.5935E-14 |
| 3s3p $^1P_1$ | 3s6g $^3G_3$ | 125.4090215 | 9.1925E-3 | 5.8022E-8 | 3.1922E-17 |
| 3s3p $^1P_1$ | 3s7d $^3D_1$ | 121.1922603 | 0.072548 | 1.0005E-5 | 2.2031E-15 |
| 3s3p $^1P_1$ | 3s7d $^3D_3$ | 121.1916141 | 0.061615 | 7.2170E-6 | 1.5891E-15 |
| 3s3p $^1P_1$ | 3s7g $^3G_3$ | 120.4825373 | 0.072796 | 4.4460E-6 | 2.2576E-15 |
| 3s3p $^1P_1$ | 3s9s $^3S_1$ | 117.9354874 | 0.49607 | 5.3606E-4 | 1.1178E-13 |
| 3s3p $^1P_1$ | 3s8d $^3D_1$ | 117.9384639 | 0.060558 | 3.4232E-6 | 1.6657E-15 |
| 3s3p $^1P_1$ | 3s8d $^3D_3$ | 117.937838 | 0.24424 | 5.5685E-5 | 2.7095E-14 |
| 3s3p $^1P_1$ | 3s8g $^3G_3$ | 117.4879939 | 0.10343 | 2.3751E-5 | 4.9150E-15 |

**Table 6**. The transition amplitudes of the M2 transitions arising from 5 lowest energy states.